\documentclass[preprint,aps,12pt,showpacs,tightenlines]{revtex4}
\usepackage{amsmath}
\usepackage{amssymb}
\usepackage{epsfig}
\usepackage{graphicx}
\begin{document}

\newcommand{\beq}{\begin{eqnarray}}
\newcommand{\eeq}{\end{eqnarray}}
\newcommand{\non}{\nonumber\\ }
\newcommand{\etap}{\eta^{(\prime)} }

\newcommand{\psl}{ p \hspace{-1.8truemm}/ }
\newcommand{\nsl}{ n \hspace{-2.2truemm}/ }
\newcommand{\vsl}{ v \hspace{-2.2truemm}/ }
\newcommand{\epsl}{\epsilon \hspace{-1.8truemm}/\,  }

\def \cpl{ Chin. Phys. Lett.  }
\def \ctp{ Commun. Theor. Phys.  }
\def \epjc{ Eur. Phys. J. C }
\def \jpg{  J. Phys. G }
\def \npb{  Nucl. Phys. B }
\def \plb{  Phys. Lett. B }
\def \prd{  Phys. Rev. D }
\def \prl{  Phys. Rev. Lett.  }

\title{Branching Ratio and CP Asymmetry of $B \to
\rho \eta^{(\prime)}$ Decays in the Perturbative QCD Approach}
\author{Xin Liu, Huisheng Wang}
\author{Zhenjun Xiao}\email{xiaozhenjun@njnu.edu.cn}
\author{Libo Guo}\email{guolb@email.njnu.edu.cn}
\affiliation{Department of Physics and Institute of Theoretical Physics ,
Nanjing Normal University, Nanjing, Jiangsu 210097, P.R.China}
\author{ Cai-Dian L\"u}
\affiliation{     CCAST (World Laboratory), P.O. Box 8730, Beijing
100080, China; }
\affiliation{Institute of High Energy Physics,
CAS, P.O. Box 918(4) Beijing 100049,   China\footnote{Mailing
address.}} 
\date{\today}
\begin{abstract}
In this paper,we calculate the branching ratios and CP-violating
asymmetries for $B^0 \to \rho^0 \eta^{(\prime)}$ and $B^+\to
\rho^+ \etap$ decays in the perturbative QCD factorization
approach. In this approach, we not only calculate the usual
factorizable contributions, but also evaluate the non-factorizable
and annihilation type contributions. Besides the current-current
operators, the contributions from the QCD and electroweak penguin
operators are also taken into account. The theoretical predictions
for the branching ratios are $Br(B^+ \to \rho^+ \eta^{(\prime)})
\approx 9 \times 10^{-6}$ and $Br(B^0 \to \rho^0 \eta^{(\prime)}) \approx
5 \times 10^{-8}$, which agree well with the measured values and
currently available experimental upper limits.
We also predict large CP-violating  asymmetries in these
decays: $A_{CP}^{dir}(\rho^\pm \eta)\approx -13 \%$,
$A_{CP}^{dir}(\rho^\pm \eta^\prime)\approx -18 \%$,
$A_{CP}^{dir}(\rho^0 \eta)\approx -41\%$, $A_{CP}^{dir}(\rho^0
\eta^\prime)\approx -27\%$, $A_{CP}^{mix}(\rho^0 \eta)\approx
+25\%$, and $A_{CP}^{mix}(\rho^0 \eta^\prime) \approx +11\%$,
which can be tested by the current or future B factory experiments.
\end{abstract}


\maketitle

\section{Introduction}

Along with the great progress in theoretical studies and
experimental measurements, the charmless two-body B meson decays
are getting more and more interesting and attracting more and more
attentions, since they provide a good place for testing the
standard model (SM), studying CP violation and searching for
possible new physics beyond the SM.

For the two-body hadronic B meson decays, the dominant theoretical
error comes from the uncertainty in evaluating the hadronic matrix
element $<M_1 M_2|O_i|B>$ where $M_1$ and $M_2$ are light final
mesons. At present, the QCD factorization (QCDF) approach
\cite{bbns,bn03b} and the perturbative  QCD (PQCD) factorization
approach \cite{lb80,cl97,li2003} are the two popular methods being
used to calculate the hadronic matrix elements. The perturbative
QCD approach has been developed earlier from the QCD
hard-scattering approach \cite{lb80}. Some elements of this
approach are also present in the QCD factorization formula of Refs.~\cite{bbns,bn03b}.
The two major differences between these two
approaches are (a) the form factors are calculable perturbatively
in PQCD approach, but taken as the input parameters extracted from
other experimental measurements in the QCDF approach; and (b) the
annihilation contributions are calculable and play an important
role in producing CP violation for the considered decay modes in
PQCD approach, but it could not be evaluated reliably in QCDF
approach. Of course, one should remember that  the assumptions
behind the PQCD approach, specifically the possibility to
calculate the form factors perturbatively, are still under
discussion~\cite{ds02}. More efforts are needed to clarify these
problems.

Up to now, many B meson hadronic decay channels have been
calculated and studied phenomenologically in both the QCDF
approach \cite{bbns,bene,du02} and in the PQCD approach
\cite{luy01,kls01,li01,kklls04,wang06}. In this paper, we would
like to calculate the branching ratios and CP asymmetries for the
$B \to \rho \eta^{(\prime)}$ decays by employing the low energy
effective Hamiltonian \cite{buras96} and the PQCD approach.
Besides the usual factorizable contributions, we here are able to
evaluate the non-factorizable and the annihilation contributions
to these decays.

Theoretically, the four $B \to \rho \eta^{(\prime)}$ decays have been studied
before in  the naive or generalized factorization approach \cite{ali98},
as well as in the QCD factorization approach \cite{du02}.
On the experimental side, the branching ratios of
$B \to \rho^+ \eta, \rho^+ \eta^\prime$ decays have been
measured \cite{babar,belle,cleo,hfag},
\beq
Br(B^+ \to \rho^+ \eta )&=& \left ( 8.1 ^{+1.7}_{-1.5}\right )\times 10^{-6},\non
Br(B^+ \to \rho^+ \eta^\prime )&=& \left ( 12.9 ^{+6.2}_{-5.5}
\pm 2.0 \right )\times 10^{-6}.
\label{eq:exp}
\eeq
For $B \to \rho^0 \eta, \rho^0 \eta^\prime$ decays,
only the experimental upper limits are available now \cite{hfag}
\beq
Br(B^0 \to \rho^0 \eta)< 1.5 \times 10^{-6}, \quad
Br(B^0 \to \rho^0 \eta^\prime)< 4.3 \times 10^{-6}.
\label{eq:ulimits}
\eeq

In $B \to \rho \etap$ decays, the $B$ meson is heavy, setting at
rest and decaying into two light mesons (i.e. $\rho$ and
$\eta^{(\prime)}$ ) with large momenta. Therefore the light final
state mesons are moving very fast in the rest frame of $B$ meson.
In this case, the short distance hard process dominates the decay
amplitude. We assume that the soft final state interaction is not
important for such decays, since there is not enough time for
light mesons to exchange soft gluons. Therefore, it makes the PQCD
reliable in calculating the $B \to \rho \eta^{(\prime)}$ decays.
With the Sudakov resummation, we can include the leading double
logarithms for all loop diagrams, in association with the soft
contribution. Unlike the usual factorization approach, the hard
part of the PQCD approach consists of six quarks rather than four.
We thus call it six-quark operators or six-quark effective theory.
Applying the six-quark effective theory to B meson decays, we need
meson wave functions for the hadronization of quarks into mesons.
All the collinear dynamics are included in the meson wave
functions.

This paper is organized as follows. In Sec.~\ref{sec:f-work},
we give a brief review for the PQCD factorization approach.
In Sec.~\ref{sec:p-c}, we calculate analytically the related Feynman
diagrams and present the various decay amplitudes for the studied decay modes.
In Sec.~\ref{sec:n-d}, we show the numerical results for the branching ratios
and CP asymmetries of $B \to \rho \eta^{(')}$ decays and comparing them with
the measured values.  The summary and some discussions are
included in the final section.

\section{ Theoretical framework}\label{sec:f-work}

The three scale PQCD factorization approach has been developed
and applied in the non-leptonic
$B$ meson decays ~\cite{cl97,li2003,lb80,luy01,kls01,li01,kklls04} for
some time. In this approach, the decay amplitude is separated into
soft, hard, and harder dynamics characterized by
different energy scales $(t, m_b, M_W)$.
It is conceptually written as the convolution,
\beq
{\cal A}(B \to M_1 M_2)\sim \int\!\! d^4k_1 d^4k_2 d^4k_3\ \mathrm{Tr}
\left [ C(t) \Phi_B(k_1) \Phi_{M_1}(k_2) \Phi_{M_2}(k_3) H(k_1,k_2,k_3, t)
\right ],
\label{eq:con1}
\eeq
where $k_i$'s are momenta of light
quarks included in each mesons, and $\mathrm{Tr}$ denotes the
trace over Dirac and color indices. $C(t)$ is the Wilson coefficient
which results from the radiative corrections at short distance. In
the above convolution, $C(t)$ includes the harder dynamics at
larger scale than $M_B$ scale and describes the evolution of local
$4$-Fermi operators from $m_W$ (the $W$ boson mass) down to
$t\sim\mathcal{O}(\sqrt{\bar{\Lambda} M_B})$ scale, where
$\bar{\Lambda}\equiv M_B -m_b$. The function $H(k_1,k_2,k_3,t)$ describes the
four quark operator and the spectator quark connected by
 a hard gluon whose $q^2$ is in the order
of $\bar{\Lambda} M_B$, and includes the
$\mathcal{O}(\sqrt{\bar{\Lambda} M_B})$ hard dynamics. Therefore,
this hard part $H$ can be perturbatively calculated. The function $\Phi_M$ is
the wave function which describes hadronization of the quark and
anti-quark to the meson $M$. While the function $H$ depends on the
processes considered, the wave function $\Phi_M$ is  independent of the specific
processes. Using the wave functions determined from other well measured
processes, one can make quantitative predictions here.

Since the b quark is rather heavy we consider the $B$ meson at
rest for simplicity. It is convenient to use light-cone coordinate
$(p^+, p^-, {\bf p}_T)$ to describe the meson's momenta,
\beq
p^\pm = \frac{1}{\sqrt{2}} (p^0 \pm p^3), \quad  and \quad {\bf
p}_T = (p^1, p^2).
\eeq
Through out this paper, we use the light-cone coordinates to write the four
momentum as $(k_1^+, k_1^-, k_1^\perp)$.
Using these coordinates the $B$ meson and
the two final state meson momenta can be written as
\beq P_1 =
\frac{M_B}{\sqrt{2}} (1,1,{\bf 0}_T), \quad P_2 =
\frac{M_B}{\sqrt{2}}(1,r_\rho^2,{\bf 0}_T), \quad P_3 =
\frac{M_B}{\sqrt{2}} (0,1,{\bf 0}_T),
\eeq
respectively, where $r_\rho=m_\rho/m_B$; the light pseudoscalar
meson masses have been neglected.

For the $B \to \rho \etap$ decays considered here, only the $\rho$
meson's longitudinal part contributes to the decays, its polar
vector is $\epsilon_L=\frac{M_B}{\sqrt{2}M_\rho}
(1,-r_\rho^2,\bf{0_T})$. Putting the light (anti-) quark momenta
in $B$, $\rho$ and $\etap$ mesons as $k_1$, $k_2$, and $k_3$,
respectively, we can choose \beq k_1 = (x_1 P_1^+,0,{\bf k}_{1T}),
\quad k_2 = (x_2 P_2^+,0,{\bf k}_{2T}), \quad k_3 = (0, x_3
P_3^-,{\bf k}_{3T}). \eeq Then, the integration over $k_1^-$,
$k_2^-$, and $k_3^+$ in eq.(\ref{eq:con1}) will lead to \beq {\cal
A}(B \to \rho \etap) &\sim &\int\!\! d x_1 d x_2 d x_3 b_1 d b_1
b_2 d b_2 b_3 d b_3 \non && \quad \mathrm{Tr} \left [ C(t)
\Phi_B(x_1,b_1) \Phi_\rho(x_2,b_2) \Phi_{\etap}(x_3, b_3) H(x_i,
b_i, t) S_t(x_i)\, e^{-S(t)} \right ], \label{eq:a2} \eeq where
$b_i$ is the conjugate space coordinate of $k_{iT}$, and $t$ is
the largest energy scale in function $H(x_i,b_i,t)$. The large
logarithms ($\ln \frac{m_W}{t}$) coming from QCD radiative corrections to
four quark operators are included in the Wilson coefficients
$C(t)$. The large double logarithms ($\ln^2 x_i$) on the
longitudinal direction are summed by the threshold resummation
~\cite{li02}, and they lead to $S_t(x_i)$ which smears the
end-point singularities on $x_i$. The last term, $e^{-S(t)}$, is
the Sudakov form factor resulting from overlap of soft and
collinear divergences, which suppresses the soft dynamics
effectively ~\cite{soft}. Thus it makes the perturbative
calculation of the hard part $H$ applicable at intermediate scale,
i.e., $M_B$ scale. We will calculate analytically the function
$H(x_i,b_i,t)$ for $B \to \rho \etap$ decays in the first order in
$\alpha_s$ expansion and give the convoluted amplitudes in next
section.

\subsection{Wilson coefficients}\label{ssec:wc}

For $B \to \rho \etap$ decays, the related weak effective Hamiltonian $H_{eff}$
can be written as \cite{buras96}
\beq
\label{eq:heff} {\cal H}_{eff} =
\frac{G_{F}} {\sqrt{2}} \, \left[ V_{ub} V_{ud}^* \left (C_1(\mu) O_1^u(\mu)
+ C_2(\mu) O_2^u(\mu) \right)
- V_{tb} V_{td}^* \, \sum_{i=3}^{10} C_{i}(\mu) \,O_i(\mu)   \right] \; .
\eeq
We specify below the operators in ${\cal H}_{eff}$ for $b \to d$ transition:
\beq
\begin{array}{llllll}
O_1^{u} & = &  \bar d_\alpha\gamma^\mu L u_\beta\cdot \bar
u_\beta\gamma_\mu L b_\alpha\ , &O_2^{u} & = &\bar
d_\alpha\gamma^\mu L u_\alpha\cdot \bar
u_\beta\gamma_\mu L b_\beta\ , \\
O_3 & = & \bar d_\alpha\gamma^\mu L b_\alpha\cdot \sum_{q'}\bar
 q_\beta'\gamma_\mu L q_\beta'\ ,   &
O_4 & = & \bar d_\alpha\gamma^\mu L b_\beta\cdot \sum_{q'}\bar
q_\beta'\gamma_\mu L q_\alpha'\ , \\
O_5 & = & \bar d_\alpha\gamma^\mu L b_\alpha\cdot \sum_{q'}\bar
q_\beta'\gamma_\mu R q_\beta'\ ,   & O_6 & = & \bar
d_\alpha\gamma^\mu L b_\beta\cdot \sum_{q'}\bar
q_\beta'\gamma_\mu R q_\alpha'\ , \\
O_7 & = & \frac{3}{2}\bar d_\alpha\gamma^\mu L b_\alpha\cdot
\sum_{q'}e_{q'}\bar q_\beta'\gamma_\mu R q_\beta'\ ,   & O_8 & = &
\frac{3}{2}\bar d_\alpha\gamma^\mu L b_\beta\cdot
\sum_{q'}e_{q'}\bar q_\beta'\gamma_\mu R q_\alpha'\ , \\
O_9 & = & \frac{3}{2}\bar d_\alpha\gamma^\mu L b_\alpha\cdot
\sum_{q'}e_{q'}\bar q_\beta'\gamma_\mu L q_\beta'\ ,   & O_{10} &
= & \frac{3}{2}\bar d_\alpha\gamma^\mu L b_\beta\cdot
\sum_{q'}e_{q'}\bar q_\beta'\gamma_\mu L q_\alpha'\ ,
\label{eq:operators}
\end{array}
\eeq where $\alpha$ and $\beta$ are the $SU(3)$ color indices; $L$
and $R$ are the left- and right-handed projection operators with
$L=(1 - \gamma_5)$, $R= (1 + \gamma_5)$. The sum over $q'$ runs
over the quark fields that are active at the scale $\mu=O(m_b)$,
i.e., $(q'\epsilon\{u,d,s,c,b\})$. The PQCD approach works well
for the leading twist approximation and leading double logarithm
summation. For the Wilson coefficients $C_i(\mu)$
($i=1,\ldots,10$), we will also use the leading order (LO)
expressions, although the next-to-leading order   calculations
already exist in the literature ~\cite{buras96}. This is the
consistent way to cancel the explicit $\mu$ dependence in the
theoretical formulae.

For the renormalization group evolution of the Wilson coefficients
 from higher scale to lower scale, we use the formulae as given in
Ref.\cite{luy01} directly. At the high $m_W$ scale, the leading
order Wilson coefficients $C_i(M_W)$ are simple and can be found
easily in Ref.\cite{buras96}.

In PQCD approach, the scale t is chosen at the maximum value of
various subprocess scales  to suppress the higher order
corrections, which may be larger or smaller than the $m_b$ scale.
In the range of $ m_b\leq t < m_W$, we will evaluate the Wilson
coefficients $C_i(t)$ at scale $t$ by using the leading logarithm
running equations, as given explicitly in Eq.(C1) of
Ref.~\cite{luy01}. In numerical calculations, we also use
$\alpha_s=4\pi/[\beta_1 \ln(t^2/{\Lambda_{QCD}^{(5)}}^2)]$ which
is the leading order expression with $\Lambda_{QCD}^{(5)}=193$MeV,
derived from $\Lambda_{QCD}^{(4)}=250$MeV. Here
$\beta_1=(33-2n_f)/12$, with the appropriate number of active
quarks $n_f$: $n_f=5$ for $ t \geq m_b$.

At a given energy scale $t=m_b=4.8$ GeV, the LO Wilson coefficients $C_i(m_b)$ as
given in Ref.~\cite{luy01} are
\beq
C_1&=& -0.2703, \quad C_2= 1.1188, \quad C_3= 0.0126, \quad C_4= -0.0270, \non
C_5&=& 0.0085, \quad C_6= -0.0326, \quad C_7= 0.0011, \quad C_8= 0.0004, \non
C_9&=& -0.0090, \quad C_{10}= 0.0022.
 \label{eq:cimb}
\eeq

In the range of $ t < m_b $, then we evaluate the Wilson coefficients $C_i(t)$
by using the $C_i(m_b)$ in Eq.~(\ref{eq:cimb}) as boundary input and the leading logarithmic
running equations as given in Appendix D of Ref.~\cite{luy01} for the case of $n_f=4$.
For the Wilson coefficient $C_2(t)$, for example, the running equation in the two different
regions can be written as
\beq
C_2(t)&=& \frac{1}{2} \left ( \eta^{-6/23} + \eta^{2/23}\right ), \ \ for \ \ m_b\leq t < m_W, \\
C_2(t)&=& \frac{1}{4} \left ( \eta^{-6/23} + \eta^{2/23}\right )
\left( \xi^{-6/25} + \xi^{12/25} \right)\non
&& + \frac{1}{4} \left ( \eta^{-6/23} - \eta^{2/23}\right )
\left( \xi^{-6/25} - \xi^{12/25} \right), \ \ for \ \ t < m_b,
\eeq
where $\eta = \alpha_S(t)/\alpha_S(m_W)$ and $\xi = \alpha_S(t)/\alpha_S(m_b)$. For the running equations
of other Wilson coefficients one can see Appendix C and D of ref.~\cite{luy01}.

\subsection{Wave Functions}\label{ssec:w-f}

In the resummation procedures, the $B$ meson is treated as a
heavy-light system. In general, the B meson light-cone matrix
element can be decomposed as ~\cite{grozin,bene}
\beq
&&\int_0^1\frac{d^4z}{(2\pi)^4}e^{i\bf{k_1}\cdot z}
   \langle 0|\bar{b}_\alpha(0)d_\beta(z)|B(p_B)\rangle \nonumber\\
&=&-\frac{i}{\sqrt{2N_c}}\left\{(\psl_B+m_B)\gamma_5
\left[\phi_B ({\bf k_1})-\frac{\nsl-\vsl}{\sqrt{2}}
\bar{\phi}_B({\bf k_1})\right]\right\}_{\beta\alpha}, \label{aa1}
\eeq
 where $n=(1,0,{\bf 0_T})$, and $v=(0,1,{\bf 0_T})$ are the
unit vectors pointing to the plus and minus directions,
respectively. From the above equation, one can see that there are
two Lorentz structures in the B meson distribution amplitudes.
They obey to the following normalization conditions
 \beq
 \int\frac{d^4 k_1}{(2\pi)^4}\phi_B({\bf
k_1})=\frac{f_B}{2\sqrt{2N_c}}, ~~~\int \frac{d^4
k_1}{(2\pi)^4}\bar{\phi}_B({\bf k_1})=0.
 \eeq

In general, one should consider these two Lorentz structures in
calculations of $B$ meson decays. However, it can be argued that
the contribution of $\bar{\phi}_B$ is numerically small ~
\cite{kurimoto}, thus its contribution can be numerically
neglected. Using this approximation, we can reduce one input
parameter in our calculation.
 Therefore, we only consider the contribution of Lorentz
structure
\beq
\Phi_B= \frac{1}{\sqrt{2N_c}} (\psl_B +m_B)
\gamma_5 \phi_B ({\bf k_1}). \label{bmeson}
\eeq
   In the next section, we
will see that the hard part is always independent of one of the
$k_1^+$ and/or $k_1^-$, if we make  approximations shown in next
section. The B meson wave function is then the function of
variable $k_1^-$ (or $k_1^+$) and $k_1^\perp$,
 \beq \phi_B (k_1^-,
k_1^\perp)=\int d k_1^+ \phi (k_1^+, k_1^-, k_1^\perp).
\label{int} \eeq

The wave function for $d\bar{d}$ components in $\eta^{(\prime)}$
meson are given as:
\beq \Phi_{\eta_{d\bar{d}}}(P,x,\zeta)\equiv
\frac{1}{\sqrt{2N_c}} \left\{\psl
\phi_{\eta_{d\bar{d}}}^{A}(x)+m_0^{\eta_{d\bar{d}}}
\phi_{\eta_{d\bar{d}}}^{P}(x)+\zeta m_0^{\eta_{d\bar{d}}} (\vsl
\nsl-v\cdot n)\phi_{\eta_{d\bar{d}}}^{T}(x)\right\}
\eeq
 where $P$ and
$x$ are the momentum and the momentum fraction of
$\eta_{d\bar{d}}$, respectively. We assumed here that the wave
function of $\eta_{d\bar{d}}$ is same as the $\pi$ wave function.
The parameter $\zeta$ is either $+1$ or $-1$ depending on the
assignment of the momentum fraction $x$.

In $B \to \rho \eta^{(\prime)}$ decays, $\rho$ meson is
longitudinally polarized. We only consider its wave function in
longitudinal polarization ~\cite{kurimoto,ball2},
 \beq
<\rho^-(P,\epsilon_L)|\bar{d_{\alpha}}(z)u_{\beta}(0)|0>=
\frac{1}{\sqrt{2N_c}}\int_0^1 d x e^{ixP\cdot z} \left\{ \epsl
\left[\psl_\rho \phi_\rho^t (x) + m_\rho \phi_\rho (x) \right]
+m_\rho \phi_\rho^s (x)\right\}. \eeq The second term in above
equation is the leading twist wave function (twist-2), while the
first and third terms are sub-leading twist (twist-3) wave
functions.

The transverse momentum $k^\perp$ is usually conveniently
converted to the $b$ parameter by Fourier transformation.
 The initial conditions of leasing twist $\phi_i(x)$,
$i=B,\rho,\eta, \eta'$, are of non-perturbative origin, satisfying
the normalization \beq
\int_0^1\phi_i(x,b=0)dx=\frac{1}{2\sqrt{6}}{f_i}\;, \label{no}
\eeq with $f_i$ the meson decay constants.

\section{Perturbative Calculations}\label{sec:p-c}

In the previous section we have discussed the wave functions and
Wilson coefficients of the amplitude in eq.(\ref{eq:con1}). In
this section, we will calculate the hard part $H(t)$. This part
involves the four quark operators and the necessary hard gluon
connecting the four quark operator and the spectator quark.  We
will show the whole amplitude for each diagram including wave
functions. Similar to the $B \to \pi \rho$ decays \cite{ly},
the eight diagrams contributing to the $B \to \rho \eta^{(\prime)}$ decays are
shown in Figure 1.
We first calculate the usual factorizable diagrams (a) and (b).
Operators $O_1$, $O_2$, $O_3$, $O_4$, $O_9$, and $O_{10}$ are
$(V-A)(V-A)$ currents, the sum of their amplitudes is given as
\beq
F_{e\rho}&=& 4 \sqrt{2} G_F \pi C_F m_B^4 \int_0^1 d x_{1}
dx_{3}\, \int_{0}^{\infty} b_1 db_1 b_3 db_3\, \phi_B(x_1,b_1)
 \non
& & \cdot \left\{ \left [(1+x_3) \phi_\rho (x_3, b_3)
+(1-2x_3)r_\rho (\phi_\rho^s (x_3, b_3)+\phi_\rho^t (x_3,b_3))
\right] \right.
 \non
 && \left. \alpha_s(t_e^1)
h_e(x_1,x_3,b_1,b_3)\exp[-S_{ab}(t_e^1)]
 \right.
  \non
&& \left. +2 r_\rho \phi_\rho^s (x_3, b_3) \alpha_s(t_e^2)
h_e(x_3,x_1,b_3,b_1)\exp[-S_{ab}(t_e^2)] \right\} \;,
\label{eq:ab}
\eeq
where  $C_F=4/3$ is a color factor. The explicit expressions of the functions $h_e^i$, the
energy scales $t_e^i$ and the Sudakov
factors $S_{ab}(t)$ Can be found in the Appendix.
In the above equation, we do not include the Wilson coefficients of the
corresponding operators, which are process dependent. They will be
shown later in this section for different decay channels. The
diagrams Fig.~1(a) and 1(b) are also the diagrams for the $B\to
\rho$ form factor $A_0^{B\to \rho}$. Therefore we can extract
$A_0^{B\to \rho}$ from Eq.~(\ref{eq:ab}).

\begin{figure}[t,b]
\vspace{-3 cm} \centerline{\epsfxsize=21 cm \epsffile{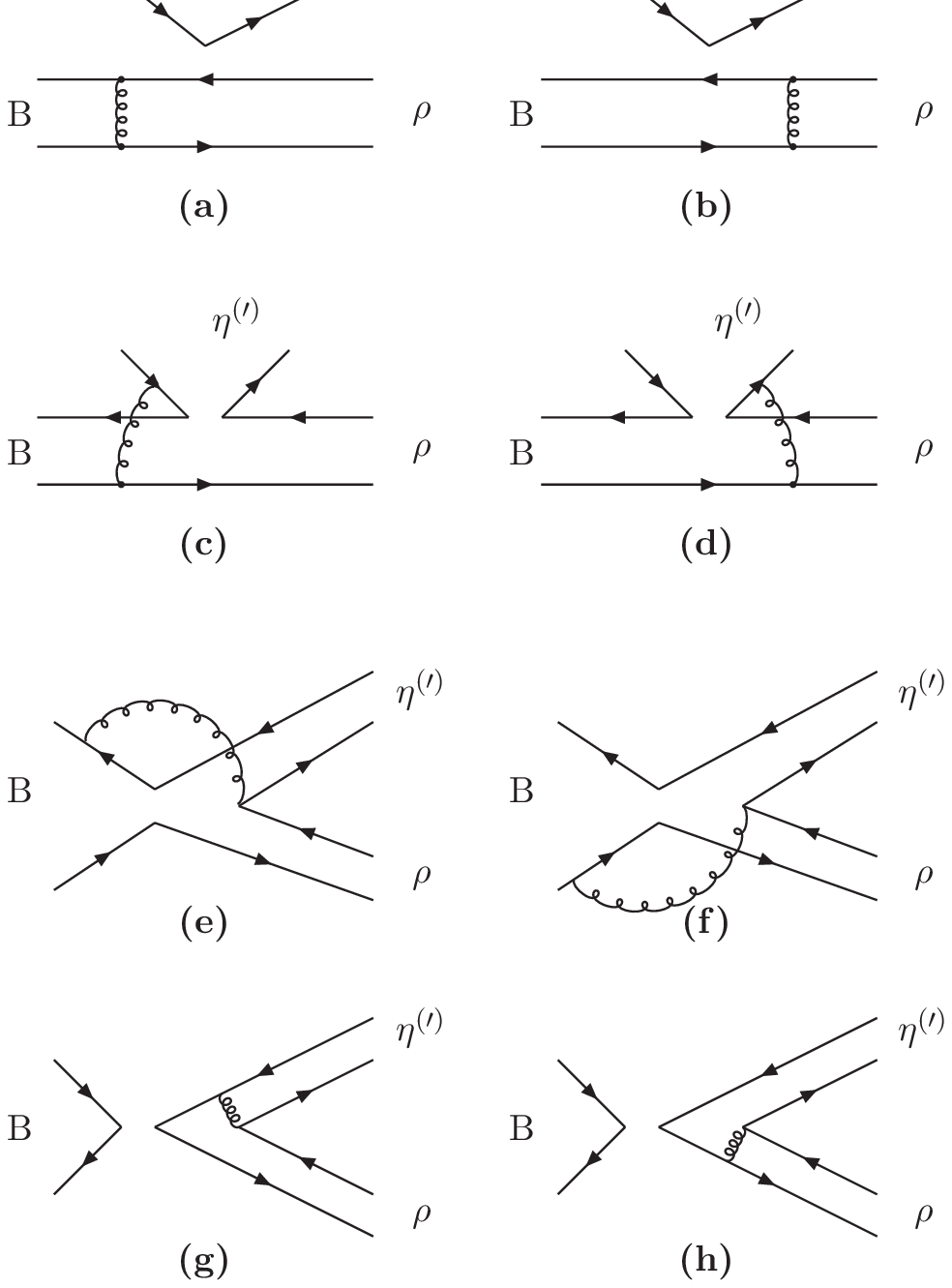}}
\vspace{-14cm} \caption{ Diagrams contributing to the $B\to
\rho\eta^{(\prime)}$
 decays (diagram (a) and (b) contribute to the $B\to \rho$ form
 factor $A_0^{B\to \rho}$).}
 \label{fig:fig1}
\end{figure}

The operators $O_5$, $O_6$, $O_7$, and $O_8$ have a structure of
$(V-A)(V+A)$. In some decay channels, some of these operators
contribute to the decay amplitude in a factorizable way. Since only
the axial-vector part of $(V+A)$ current contribute to the
pseudo-scaler meson production,
 \beq
\langle \rho |V-A|B\rangle \langle \eta |V+A | 0 \rangle =
-\langle \rho |V-A |B  \rangle \langle \eta |V-A|0 \rangle  ,
 \eeq
 the result of these operators is opposite to Eq.~(\ref{eq:ab}).
 In some other cases, we need to do Fierz
transformation  for these operators to get right color structure
for factorization to work. In this case, we get $(S+P)(S-P)$
operators from $(V-A)(V+A)$ ones. For these $(S+P)(S-P)$
operators, Fig.~ 1(a) and 1(b) give
\beq
 F_{e\rho}^{P}&=& 8 \sqrt{2} G_F \pi C_F f_\eta^d m_B^4 \int_{0}^{1}d x_{1}d
x_{3}\,\int_{0}^{\infty} b_1d b_1 b_3d b_3\, \phi_B(x_1,b_1) \non
& & \cdot
 \left\{ \left[ \phi_\rho (x_3, b_3)+r_\rho ((x_3+2) \phi_\rho^s (x_3, b_3)- x_3 \phi_\rho^t (x_3,
 b_3))\right]\right.
\non
 & &\left. \cdot \alpha_s (t_e^1)  h_e
(x_1,x_3,b_1,b_3)\exp[-S_{ab}(t_e^1)]
  \right.  \non
& &\left.  + \left(x_1 \phi_\rho(x_3,b_3)+ 2 r_\rho \phi_\rho^s
(x_3, b_3)\right) \alpha_s (t_e^2)
 h_e(x_3,x_1,b_3,b_1)\exp[-S_{ab}(t_e^2)] \right\} \;.
\eeq

For the non-factorizable diagrams (c) and (d), all three meson
wave functions are involved. The integration of $b_3$ can be
performed using $\delta$ function $\delta(b_3-b_2)$, leaving only
integration of $b_1$ and $b_2$.
For the $(V-A)(V-A)$ operators, the result is
\beq
 M_{e\rho}&=& - \frac{16} {\sqrt{3}} G_F \pi C_F m_B^4
\int_{0}^{1}d x_{1}d x_{2}\,d x_{3}\,\int_{0}^{\infty} b_1d b_1
b_2d b_2\, \phi_B(x_1,b_1) \phi_\eta^A(x_2,b_2) \non
 & &\cdot
\left \{ x_3 \left[\phi_\rho(x_3,b_2)-2 r_\rho
\phi_\rho^t(x_3,b_3)\right]
 \alpha_s(t_f)
 h_f(x_1,x_2,x_3,b_1,b_2)\exp[-S_{cd}(t_f)]\right \} \; .
\eeq

For the non-factorizable annihilation diagrams (e) and (f), again
all three wave functions are involved. Here we have two kinds of
contributions. $M_{a\rho}$ is the contribution containing operator
type $(V-A)(V-A)$, while $M_{a\rho}^{P}$ is the contribution
containing operator type $(V-A)(V+A)$.
 \beq
M_{a\rho}&=& \frac{16} {\sqrt{3}} G_F \pi C_F m_B^4 \int_{0}^{1}d
x_{1}d x_{2}\,d x_{3}\,\int_{0}^{\infty} b_1d b_1 b_2d b_2\,
\phi_B(x_1,b_1) \non && \cdot \left\{ \left[ x_3 \phi_\rho(x_3,
b_2) \phi_\eta^A(x_2,b_2) + r_\rho r_\eta \left( (x_3-x_2)
\left(\phi_\eta^P(x_2,b_2) \phi_\rho^t (x_3,b_2)+\phi_\eta^T
(x_2,b_2) \right.\right.\right.\right.
 \non
 &&\left. \left.\left.\left. \cdot \phi_\rho^s(x_3,b_2) \right)+(x_3+x_2)
 \left(\phi_\eta^P(x_2,b_2) \phi_\rho^s(x_3,b_2)
   +\phi_\eta^T (x_2,b_2) \phi_\rho^t(x_3,b_2)\right)\right)\right]\right.
   \non
&&\left.\cdot  \alpha_s(t_f^1)
h_f^1(x_1,x_2,x_3,b_1,b_2)\exp[-S_{ef}(t_f^1)]~-~\left[x_2
\phi_\rho(x_3,b_2) \phi_\eta^A(x_2,b_2)\right.\right. \non
 &&\left.\left.
 + r_\rho r_\eta \left( (x_2-x_3)\left(\phi_\eta^P(x_2,b_2)\phi_\rho^t(x_3,b_2)
 +\phi_\eta^T(x_2,b_2)\phi_\rho^s(x_3,b_2)\right)~+~r_\rho r_\eta
 \right.\right.\right.
\non && \left.\left.\left.\cdot
\left((2+x_2+x_3)\phi_\eta^P(x_2,b_2) \phi_\rho^s(x_3,b_2)
  - (2-x_2-x_3)\phi_\eta^T(x_2,b_2) \phi_\rho^t(x_3,b_2)\right)\right) \right]\right.
 \non
 & &\left. \cdot
 \alpha_s(t_f^2)h_f^2(x_1,x_2,x_3,b_1,b_2)\exp[-S_{ef}(t_f^2)] \right\}
 \;,
\eeq
where $r_\eta \equiv r_\pi= m_0^{\pi}/m_B $.
 \beq
 M_{a\rho}^{P}&=& -\frac{16} {\sqrt{3}} G_F \pi C_F
m_B^4 \int_{0}^{1}d x_{1}d x_{2}\,d x_{3}\,\int_{0}^{\infty} b_1d
b_1 b_2d b_2\, \phi_B(x_1,b_1)
 \non
& &\cdot \left\{ \left[x_2 r_\eta \phi_\rho(x_3, b_2) \left
(\phi_\eta^P(x_2,b_2)+\phi_\eta^T(x_2,b_2) \right) - x_3 r_\rho
\left(\phi_\rho^s(x_3,b_2)+\phi_\rho^t(x_3,b_2) \right)
\right.\right.
 \non
 & &\left. \cdot \phi_\eta^A(x_2,b_2)]\alpha_s(t_f^1)h_f^1(x_1,x_2,x_3,b_1,b_2)
\exp[-S_{ef}(t_f^1)]\right.
 \non
& &\left.+\left[(2-x_2)r_\eta \phi_\rho(x_3, b_2) \left
(\phi_\eta^P(x_2,b_2)+\phi_\eta^T(x_2,b_2) \right)- (2-x_3)
r_\rho\left(\phi_\rho^s(x_3,b_2) \right.\right.\right.
 \non
 & &\left.\left.\left.
+\phi_\rho^t(x_3,b_2) \right)
 \phi_\eta^A(x_2,b_2) \right]\alpha_s(t_f^2)h_f^2(x_1,x_2,x_3,b_1,b_2)\exp[-S_{ef}(t_f^2)]\}\;.
\right.
\eeq

The factorizable annihilation diagrams (g) and (h) involve only
$\rho$ and $\etap$ wave functions. There are also two kinds of
decay amplitudes for these two diagrams. $F_{a\rho}$ is for
$(V-A)(V-A)$ type operators, and $F_{a\rho}^{P}$ is for
$(S-P)(S+P)$ type operators,
 \beq
 F_{a\rho}&=& -4 \sqrt{2} \pi G_F C_F f_B m_B^4 \int_{0}^{1}d
x_{2}\,d x_{3}\,\int_{0}^{\infty} b_2d b_2b_3d b_3 \,
 \non
& &\cdot \left\{ \left[ x_3 \phi_\rho(x_3,b_3)
\phi_\eta^A(x_2,b_2) + 2 r_\rho r_\eta \phi_\eta^P(x_2,b_2)
\left((1+x_3)\phi_\rho^s(x_3, b_3)\right.\right.\right. \non
 &&\left.\left.\left.-(1-x_3) \phi_\rho^t(x_3,b_2) \right)
\right]\alpha_s(t_e^3)
h_a(x_2,x_3,b_2,b_3)\exp[-S_{gh}(t_e^3)]\right.
\non
 &&\left. +\left[x_2 \phi_\rho(x_3,b_3) \phi_\eta^A(x_2,b_2)+2 r_\rho
r_\eta
\phi_\rho^s(x_3,b_3)\left((1+x_2)\phi_\eta^P(x_2,b_2)\right.\right.\right.
 \non
& &\left.\left.\left.-(1-x_2)\phi_\eta^T(x_2,b_2) \right) \right]
 \alpha_s(t_e^4)
 h_a(x_3,x_2,b_3,b_2)\exp[-S_{gh}(t_e^4)]\right \}\;,
\eeq
\beq
 F_{a\rho}^{P}
&=& -8 \sqrt{2} G_F \pi C_F  m_B^4 f_B \int_{0}^{1}d x_{2}\,d
x_{3}\,\int_{0}^{\infty} b_2d b_2b_3d b_3 \,
 \non
& &\cdot \left\{ \left[2 r_\eta \phi_\rho(x_3, b_3)
\phi_\eta^P(x_2,b_2) + x_3 r_\rho \left(\phi_\rho^s(x_3, b_3)-
\phi_\rho^t(x_3,b_2) \right) \phi_\eta^A(x_2,b_2) \right]\right.
 \non
&&\left.\cdot \alpha_s(t_e^3)
h_a(x_2,x_3,b_2,b_3)\exp[-S_{gh}(t_e^3)]\right.
 \non
 &&\left.+\left[2 r_\rho
\phi_\rho^s(x_3,b_3) \phi_\eta^A(x_2,b_2)+ x_2 r_\eta
 \left
(\phi_\eta^P(x_2,b_2)-\phi_\eta^T(x_2,b_2)\right)\phi_\rho(x_3,b_3)
\right]\right.
 \non
 &&\left. \cdot
 \alpha_s(t_e^4)
 h_a(x_3,x_2,b_3,b_2)\exp[-S_{gh}(t_e^4)]\right \}
\;.
\eeq

In the above equations, we have assumed that $x_1 <<x_2,x_3$.
Since the light quark momentum fraction $x_1$ in $B$ meson is
peaked at the small $x_1$ region, while quark momentum fraction
$x_2$ of $\etap$ is peaked around $0.5$, this is not a bad
approximation. The numerical results also show that this
approximation makes very little difference in the final result.
After using this approximation, all the diagrams are functions of
$k_1^-= x_1 m_B/\sqrt{2}$ of B meson only, independent of the
variable of $k_1^+$. Therefore the integration of eq.(\ref{int})
is performed safely.

If we exchange the $\rho$ and $\etap$ in Figure 1, the result will
be different. Because this will switch the dominant contribution
 from $B \to \rho$ form factor to $B \to \etap$ form factors. The new
diagrams are shown in Figure 2.

\begin{figure}[t,b]
\vspace{-3 cm} \centerline{\epsfxsize=21 cm \epsffile{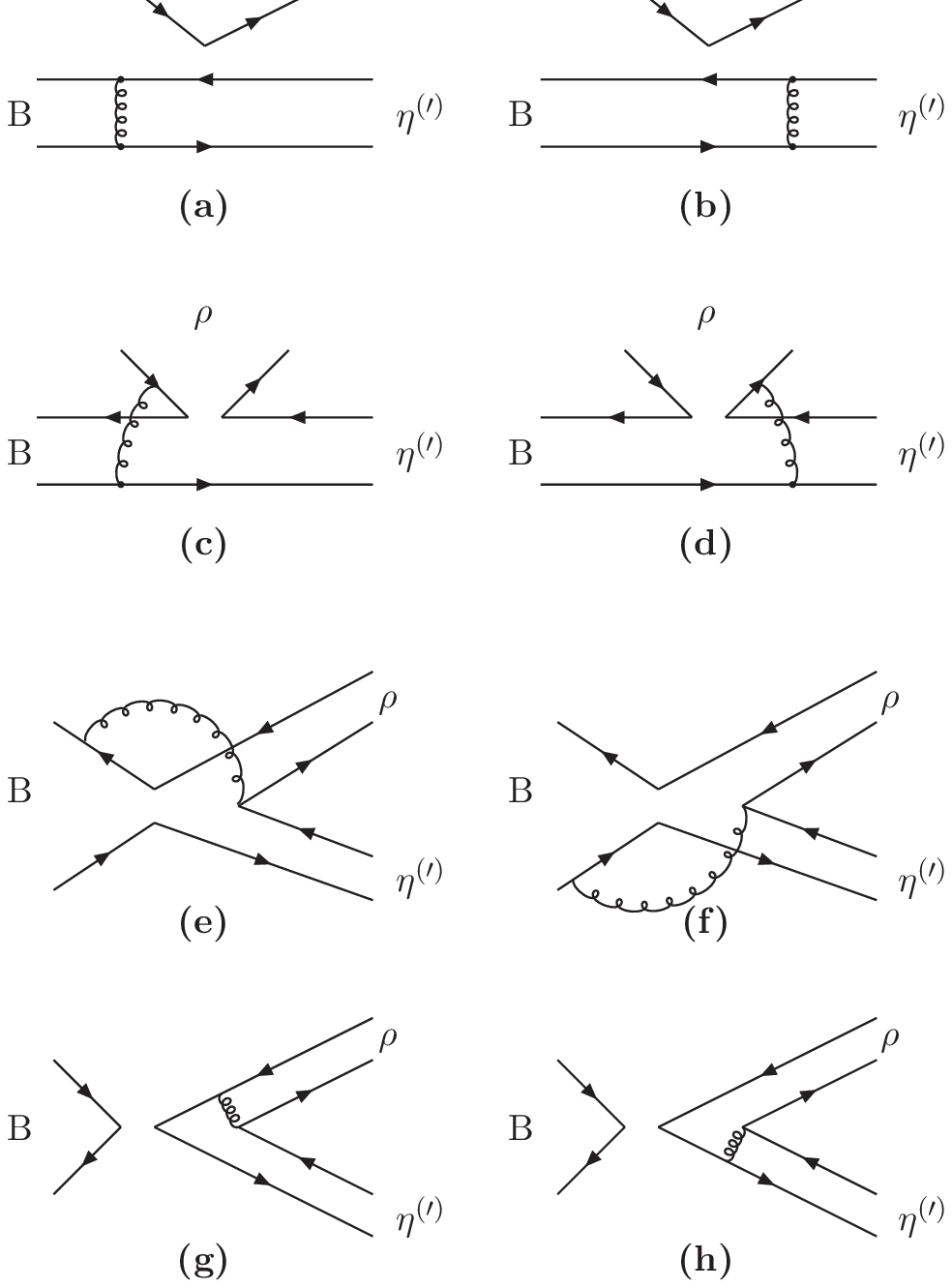}}
\vspace{-14cm} \caption{ Diagrams contributing to the $B\to
\rho\etap$
 decays (diagram (a) and (b) contribute to the $B\to \etap$ form
 factor $F_0^{B\to \etap}$).}
 \label{fig:fig2}
\end{figure}

We firstly consider the factorizable diagrams Fig.~2(a) and 2(b).
The decay amplitude $F_e$ induced by inserting the $(V-A)(V-A)$
operators is
\beq
F_e&=& 4 \sqrt{2} \pi G_F C_F f_\rho m_B^4
\int_{0}^{1}d x_{1}d x_{3}\,\int_{0}^{\infty} b_1d b_1 b_3d b_3\,
\phi_B(x_1,b_1)
 \non
& & \cdot \left\{ \left[(1+x_3)\phi_\eta^A(x_3,b_3)
 + r_\eta  (1-2x_3) \left(\phi_\eta^P(x_3,b_3) +\phi_\eta^T(x_3,b_3)
 \right)
 \right] \right.
 \non
& & \left. \cdot \alpha_s (t_e^1)
h_e(x_1,x_3,b_1,b_3)\exp[-S_{ab}(t_e^1)]\right.
 \non
& & \left. + 2 r_\eta \phi_\eta^P(x_3,b_3) \alpha_s (t_e^2)
h_e(x_3,x_1,b_3,b_1)\exp[-S_{ab}(t_e^2)] \right \}\;.\label{ls}
\label{eq:fe2} \eeq
These two diagrams are also responsible for
the calculation of $B \to \etap$ form factors $F_0^{B\to \eta}$
and $F_0^{B\to \eta^\prime}$, These two form factors can be
extracted from Eq.~(\ref{ls}).

Since only the vector part of the $(V+A)$ current contribute to
the vector meson production, the decay amplitude $F_e^{P}$ induced
by inserting $(V-A)(V+A)$ operators is identical with the
amplitude $F_e$ as given in Eq.~(\ref{eq:fe2}), i.e.,
\beq
F_e^{P}&=&F_e .
\eeq

Because neither scaler nor pseudo-scaler density gives
contribution to a vector meson production, $\langle \rho|S+P|
0\rangle =0$, we get $F_e^{S+P}=  0$.

For the non-factorizable diagrams Fig.~2(c) and 2(d), the
corresponding decay amplitudes are
\beq
 M_e&=& -\frac{16}
{\sqrt{3}} G_F \pi C_F  m_B^4 \int_{0}^{1}d x_{1}d x_{2}\,d
x_{3}\,\int_{0}^{\infty} b_1d b_1 b_2d b_2\, \phi_B(x_1,b_1)
\phi_\rho(x_2,b_2)
 \non
& &\cdot \left\{x_3 \left[ \phi_\eta^A(x_3, b_2)- 2 r_\eta
\phi_\eta^T(x_3,b_2) \right] \alpha_s(t_f)
h_f(x_1,x_2,x_3,b_1,b_2)\exp[-S_{cd}(t_f)]\right \} \;,
 \eeq
\beq M_e^{P}&=& - \frac{32} {\sqrt{3}} G_F \pi C_F r_\rho m_B^4
\int_{0}^{1}d x_{1}d x_{2}\,d x_{3}\,\int_{0}^{\infty} b_1d b_1
b_2d b_2\, \phi_B(x_1,b_1) \non & &\cdot \left\{ \left[ x_2
\phi_\eta^A(x_3, b_2)
\left(\phi_\rho^s(x_2,b_2)-\phi_\rho^t(x_2,b_2)\right) + r_\eta
\left( (x_2+x_3)
\left(\phi_\eta^P(x_3,b_2)\right.\right.\right.\right. \non
 && \left.\left.\left.\left.\cdot \phi_\rho^s(x_2,b_2)
 + \phi_\eta^T(x_3,b_2)\phi_\rho^t(x_2,b_2)\right)
 +(x_3-x_2)\left(\phi_\eta^P(x_3,b_2)\phi_\rho^t(x_2,b_2)\right.\right.\right.\right.
 \non
 && \left.\left.\left.\left.+\phi_\eta^T(x_3,b_2)\phi_\rho^s(x_2,b_2)\right)\right)\right]
 \alpha_s(t_f)h_f(x_1,x_2,x_3,b_1,b_2)\exp[-S_{cd}(t_f)]\right \}
\;. \eeq

 From  the non-factorizable annihilation diagrams Fig.~2(e) and
2(f), we find the decay amplitude $M_a$ for $(V-A)(V-A)$
operators, $M_a^{P}$ for $(V-A)(V+A)$ operators,
 \beq
 M_a&=& \frac{16}
{\sqrt{3}} \pi G_F C_F  m_B^4 \int_{0}^{1}d x_{1}d x_{2}\,d
x_{3}\,\int_{0}^{\infty} b_1d b_1 b_2d b_2\, \phi_B(x_1,b_1)
 \non
 & &\cdot \left\{ \left[x_3 \phi_\rho(x_2,b_2)\phi_\eta^A(x_3,b_2)+r_\rho
r_\eta
\left((x_3-x_2)\left(\phi_\eta^P(x_3,b_2)\phi_\rho^t(x_2,b_2)+
\phi_\eta^T(x_3,b_2)\right.\right.\right.\right.
 \non
 &&
 \left.\left.\left.\left.\cdot \phi_\rho^s(x_2,b_2)\right)+(x_2+x_3)
 \left(\phi_\eta^P(x_3,b_2)\phi_\rho^s(x_2,b_2)+
\phi_\eta^T(x_3,b_2)\phi_\rho^t(x_2,b_2)\right) \right)
\right]\right.
 \non
 &&\left. \cdot \alpha_s(t_f^1)
h_f^1(x_1,x_2,x_3,b_1,b_2)\exp[-S_{ef}(t_f^1)]\right.
 \non
&&\left. +\left[x_2 \phi_\rho(x_2,b_2)\phi_\eta^A(x_3, b_2)+r_\rho
 r_\eta\left((x_2-x_3)\left(\phi_\eta^P(x_3,b_2) \phi_\rho^t(x_2,b_2)
 +\phi_\eta^T(x_3,b_2)\right.\right.\right.\right.
 \non
  && \left.\left.\left.\left.\cdot
 \phi_\rho^s(x_2,b_2)\right)+(2+x_2+x_3)\phi_\eta^P(x_3,b_2)\phi_\rho^s(x_2,b_2)
 -(2-x_2-x_3)\phi_\eta^T(x_3,b_2)\right.\right.\right.
 \non
 &&\left.\left.\left.\cdot \phi_\rho^t(x_2,b_2) \right)
\right]
 \alpha_s(t_f^2)
 h_f^2(x_1,x_2,x_3,b_1,b_2)\exp[-S_{ef}(t_f^2)]\right \}\; ,
 \eeq
\beq
 M_a^{P} &=& M_{a\rho}^{P}.
\eeq

For the factorizable annihilation diagrams Fig.~2(g) and 2(h), we
have
\beq
 F_a&=& -F_{a\rho}, \ \  and  \  \ F_a^{P} = -F_{a\rho}^{P}.
\eeq

Now we are able to calculate perturbatively the form factors
$F_0^{B\to \eta^{(')}}(0)$, $A_{0,1}^{B\to \rho} (0)$,
and the decay amplitudes for the Feynman diagrams after the integration over $x_i$ and $b_i$.
When doing the above integrations over $x_i$ and $b_i$, we should include
the corresponding Wilson coefficients $C_i(t_j)$ calculated at the appropriate scale $t_j$ using
Eqs.~(C1) and (D1) of Ref.~\cite{luy01}.
Since we here calculated the form factors and amplitudes
at the leading order ( one order of $\alpha_s(t)$), the radiative corrections at the next order would emerge
in terms of $\alpha_s(t) \ln(m/t)$, where $m'$s denote some scales, like $m_B, 1/b_i, \ldots$,
in the hard part $H(t)$. We select the largest energy scale among $m'$s
appearing in each diagram as the hard scale $t'$s for the purpose of at least  killing
the large logarithmic corrections partially,
\beq
t_{e}^1 &=& a_t \cdot {\rm max}(\sqrt{x_3} m_B,1/b_1,1/b_3)\;,\label{t1}\non
t_{e}^2 &=& a_t \cdot {\rm max}(\sqrt{x_1}m_B,1/b_1,1/b_3)\;,\non
t_{e}^3 &=& a_t \cdot {\rm max}(\sqrt{x_3}m_B,1/b_2,1/b_3)\;,\non
t_{e}^4 &=& a_t \cdot {\rm max}(\sqrt{x_2}m_B,1/b_2,1/b_3)\;,\non
t_{f}   &=&   a_t \cdot {\rm max}(\sqrt{x_1 x_3}m_B, \sqrt{x_2 x_3} m_B,1/b_1,1/b_2)\;,\non
t_{f}^1 &=& a_t \cdot {\rm max}(\sqrt{x_2 x_3} m_B,1/b_1,1/b_2)\;,\non
 t_{f}^2 &=& a_t \cdot {\rm max}(\sqrt{x_1+x_2+x_3-x_1 x_3-x_2 x_3}m_B, \sqrt{x_2 x_3} m_B,1/b_1,1/b_2)\;,
\label{tf}
\eeq
where the constant $a_t=1\pm 0.1$ is introduced in order to estimate the scale dependence of the theoretical
predictions for the observables.

Before we put the things together to write down the decay amplitudes for the
studied decay modes, we give a brief discussion about the $\eta-\eta^\prime$ mixing
and the gluonic component of the $\eta^\prime$ meson.

The $\eta$ and $\eta^\prime$  are neutral pseudoscalar ($J^P=0^-$) mesons, and usually
considered as mixtures of the $SU(3)_F$ singlet $\eta_1$ and the octet $\eta_8$:
\beq
\left(\begin{array}{c} \eta \\ \eta^{\prime} \end{array} \right)
= \left(\begin{array}{cc}
 \cos{\theta_p} & -\sin{\theta_p} \\
 \sin{\theta_p} & \cos{\theta_p} \\ \end{array} \right)
 \left(\begin{array}{c}
 \eta_8 \\ \eta_1 \end{array} \right),
\label{eq:e-ep}
\eeq
with
\beq
\eta_8&=&\frac{1}{\sqrt{6}}\left ( u\bar{u}+d\bar{d}-2s\bar{s}\right ),\non
\eta_1&=&\frac{1}{\sqrt{3}}\left (u\bar{u}+d\bar{d}+s\bar{s}\right ),
\label{eq:e1-e8}
\eeq
where $\theta_p$ is the mixing angle to be determined by various related
experiments \cite{pdg04}. From previous studies, one obtains the mixing angle
$\theta_p$ between $-20^{\circ}$ to $-10^{\circ}$.
One best fit result as given in Ref.~\cite{ekou01} is
$-17^{\circ}\leq \theta_p \leq -10^{\circ}$.

As shown in Eqs.~(\ref{eq:e-ep},\ref{eq:e1-e8}),  $\eta$ and
$\eta^\prime$ are generally considered as a linear combination of
light quark pairs. But it should be noted that the $\eta^\prime$
meson may has a gluonic component in order to interpret the
anomalously large branching ratios of $B\to K \eta^\prime$ and
$J/\Psi \to \eta^\prime \gamma$ \cite{ekou01,ekou02}. In
Refs.~\cite{rosner83,ekou01,ekou02}, the physical states $\eta$
and $\eta^\prime$ were defined as
\beq |\eta> &=& X_\eta \left |
\frac{u\bar{u} + \bar{d}d}{\sqrt{2}} \right > + Y_\eta | s
\bar{s}>, \non |\eta^\prime> &=& X_{\eta^\prime} \left |
\frac{u\bar{u} + \bar{d}d}{\sqrt{2}} \right > + Y_{\eta^\prime} |
s \bar{s}> + Z_{\eta^\prime} |gluonium>, \non
\eeq
where $X_{\etap}, Y_{\etap}$ and $Z_{\eta'}$ parameters describe the
ratios of $u\bar{u}+d \bar{d}$, $s \bar{s}$ and gluonium
($SU(3)_F$ singlet) component of $\etap$, respectively. In
Ref.\cite{ekou01}, the author shows that the gluonic admixture in
$\eta^\prime$ can be as large as $26\%$, i.e.
\beq
Z_{\eta'}/\left ( X_{\etap}+Y_{\etap}+Z_{\eta'}\right ) \leq 0.26.
\eeq
According to paper \cite{ekou02}, a large SU(3) singlet contribution can
help us to explain the large branching ratio for $B \to K
\eta^\prime$ decay, but also result in a large branching ratio for
$B \to K^0 \eta$ decay, $Br(B \to K^0 \eta) \sim 7.0 (13) \times
10^{-6}$ for $\theta_P =-20^\circ (-10^\circ)$ as given in Table
II of Ref.~\cite{ekou02}, which is clearly too large than
currently available upper limits \cite{hfag}:
\beq
Br(B \to K^0 \eta) < 1.9  \times 10^{-6}.
\eeq

Although a lot of studies have been done along this direction, but we currently still
do not understand the anomalous  $gg-\eta^\prime$ coupling clearly, and do not know
how to calculate reliably the contributions induced by the gluonic component of
$\eta^\prime$ meson. In this paper, we firstly assume that $\eta^\prime$ does not have
the gluonic component,  and set the quark content of $\eta$ and $\eta^\prime$
as described by Eqs.~(\ref{eq:e-ep},\ref{eq:e1-e8}).
We will also discuss the effects of a non-zero gluonic admixture of $\eta^\prime$ in
next section.


Combining the contributions from different diagrams, the total
decay amplitude for $B^+ \to \rho^+ \eta$ decay can be written as
\beq
\sqrt{3} {\cal M}(\rho^+ \eta) &=& F_{e\rho} \left \{\left[
\xi_u \left( C_1 + \frac{1}{3}C_2\right)-\xi_t
\left(-\frac{1}{3}C_3-C_4-\frac{3}{2}C_7-\frac{1}{2}C_8+\frac{5}{3}C_9
\right.\right.\right.
 \non
& &\left.\left.\left. + C_{10}\right)\right ] f_\eta^d F_1(\theta_p)~-~
\xi_t\left(\frac{1}{2}C_7+\frac{1}{6}C_8-\frac{1}{2}C_9-\frac{1}{6}C_{10}\right)
f_\eta^s F_2(\theta_p)\right \}
 \non
&& - F_{e\rho}^{P} \xi_t \left (\frac{1}{3}C_5+C_6
-\frac{1}{6}C_7-\frac{1}{2}C_{8}\right) F_1(\theta_p) \non && +
M_{e\rho}\left \{ \left [ \xi_uC_2-\xi_t \cdot
\left(C_3+2C_4 +2C_6+\frac{1}{2}C_8 -\frac{1}{2}C_9
+\frac{1}{2}C_{10}\right)\right ] F_1(\theta_p) \right. \non &&
\left. -\xi_t \left ( C_4+C_6
-\frac{1}{2}C_8-\frac{1}{2}C_{10}\right ) F_2(\theta_p) \right \}
\non && +(M_{a\rho}+M_e+M_a)\left [\xi_u C_1 - \xi_t ( C_3+C_9
)\right  ] \cdot F_1(\theta_p)\non && -\left (2
M_{a\rho}^{P}+M_e^{P} \right ) \, \xi_t \,(C_5+
 C_7 ) \cdot  F_1(\theta_p)
 \non
 && + F_e \left \{ \left [\xi_u \left(\frac{1}{3}C_1+C_2\right)-
\xi_t \left(\frac{1}{3}C_3+ C_4 +\frac{1}{3}C_9 + C_{10}
\right)\right ] F_1(\theta_p) \right \} ,
\label{eq:m1}
\eeq
where $\xi_u = V_{ub}^*V_{ud}$, $\xi_t = V_{tb}^*V_{td}$, and
$F_1(\theta_p)= -\sin \theta_p + \cos \theta_p/\sqrt{2}$ and
$F_2(\theta_p)=-\sin \theta_p -\sqrt{2}\cos \theta_p$  are the
mixing factors. The Wilson coefficients $C_i$ should be calculated
at the appropriate scale $t$ using equations as given in the
Appendices of Ref.\cite{luy01}.

Similarly, the decay amplitude for $B^0 \to \rho^0 \eta$ can be written as
\beq
\sqrt{6}{\cal M}(\rho^0 \eta) &=& F_e \left [ \xi_u \left(C_1+
\frac{1}{3}C_2\right) - \xi_t \left (-\frac{1}{3}C_3
-C_4+\frac{3}{2}C_7+\frac{1}{2}C_8+\frac{5}{3}C_9
+C_{10}\right)\right ] F_1(\theta_p) \non
&&
- F_{e\rho}\left \{ \left [ \xi_u\left (C_1+\frac{1}{3}C_2 \right )\right.\right.\non
&& \left.\left.
 -\xi_t\left(\frac{1}{3}C_3+C_4
-\frac{1}{2}C_7-\frac{1}{6}C_8+\frac{1}{3}C_9-\frac{1}{3}
C_{10}\right)\right] f_\eta^d F_1(\theta_p) \right. \non && \left.
+\xi_t\left(\frac{1}{2}C_7+\frac{1}{6}C_8
+\frac{1}{2}C_9+\frac{1}{6}C_{10}\right) f_\eta^s  F_2(\theta_p)
\right \} \non && + F_{e\rho}^{P}  \; \xi_t \left(\frac{1}{3}C_5+
C_6-\frac{1}{6}C_7 -\frac{1}{2}C_8 \right) \cdot F_1(\theta_p)
\non && - M_{e\rho}\left \{ \left [\xi_u C_2 -\xi_t \left(C_3 + 2
C_4+2C_6+\frac{1}{2}C_8-\frac{1}{2}C_9
+\frac{1}{2}C_{10}\right)\right ] \cdot F_1(\theta_p)\right. \non
&& \left. -\xi_t \left(
C_4+C_6-\frac{1}{2}C_8-\frac{1}{2}C_{10}\right )
F_2(\theta_p)\right \}\non && +\left (M_{a\rho}+M_a\right) \left
[\xi_u C_2-\xi_t\left( -C_3+\frac{3}{2}C_8+\frac{1}{2}C_9
+\frac{3}{2}C_{10}\right)\right ] F_1(\theta_p)\non
 && - \left(M_e^{P}+2 M_a^{P}\right)
 \; \xi_t \; \left (C_5-\frac{1}{2}C_7 \right )
 F_1(\theta_p)  \non
&& + M_e \left[ \xi_u C_2 -\xi_t \left(-C_3 -\frac{3}{2}C_8+\frac{1}{2}C_9
+\frac{3}{2}C_{10}\right)\right ] F_1(\theta_p)  .
\label{eq:m2}
\eeq

The decay amplitudes for $B \to \rho^+ \eta^{\prime}$ and $B \to \rho^0 \eta'$
can be obtained easily from Eqs.(\ref{eq:m1}) and (\ref{eq:m2}) by the following
replacements
\beq
f_\eta^{d}, f_\eta^s &\longrightarrow& f_{\eta^\prime}^d, f_{\eta^\prime}^s, \non
F_1(\theta_p) &\longrightarrow & F'_1(\theta_p)
= \cos{\theta_p} + \frac{\sin{\theta_p}}{\sqrt{2}}, \non
F_2(\theta_p) &\longrightarrow & F'_2(\theta_p) = \cos{\theta_p}
- \sqrt{2} \sin{\theta_p}.
\eeq
Note that the possible gluonic component of $\eta'$ meson has been neglected here.

\section{Numerical results and Discussions}\label{sec:n-d}

\subsection{Input parameters and wave functions}

We use the following input parameters in the numerical calculations
\beq
 \Lambda_{\overline{\mathrm{MS}}}^{(f=4)} &=& 250 {\rm MeV}, \quad
 f_\pi = 130 {\rm MeV}, f_B = 190 {\rm MeV}, \non
 m_0^{\eta_{d\bar{d}}}&=& 1.4 {\rm GeV}, \quad f_\rho = 200 {\rm MeV},
 \quad f_K = 160  {\rm MeV}, \non
 M_B &=& 5.2792 {\rm GeV}, \quad M_W = 80.41{\rm GeV}.
 \label{para}
\eeq The central values of the CKM matrix elements to be used in
numerical calculations are \cite{pdg04}
\beq |V_{ud}|&=&0.9745,
\quad |V_{ub}|=0.0040,\non |V_{tb}|&=&0.9990, \quad
|V_{td}|=0.0075.
\eeq

For the $B$ meson wave function, we adopt the model
\beq
\phi_B(x,b) &=&  N_B x^2(1-x)^2 \mathrm{exp} \left
 [ -\frac{M_B^2\ x^2}{2 \omega_{b}^2} -\frac{1}{2} (\omega_{b} b)^2\right],
 \label{phib}
\eeq
where $\omega_{b}$ is a free parameter and we take
$\omega_{b}=0.4\pm 0.04$ GeV in numerical calculations, and
$N_B=91.745$ is the normalization factor for $\omega_{b}=0.4$.
This is the same wave functions as in
Refs.~\cite{luy01,kurimoto,kls01,cl00}, which is a best fit for
most of the measured hadronic B decays.

For the light meson wave function, we neglect the $b$ dependant
part, which is not important in numerical analysis. We choose the
wave function of $\rho$ meson similar to the pion case
~\cite{ball2}
\beq
\phi_\rho(x) &=& \frac{3}{\sqrt{6} }
 f_\rho  x (1-x)  \left[1+ 0.18C_2^{3/2} (2x-1) \right],\\
    \phi_\rho^t(x) &=&  \frac{f_\rho^T }{2\sqrt{6} }
  \left\{  3 (2 x-1)^2 +0.3(2 x-1)^2  \left[5(2 x-1)^2-3  \right]
  \right.
  \nonumber\\
 &&~~\left. +0.21 [3- 30 (2 x-1)^2 +35 (2 x-1)^4] \right\},\\
\phi_\rho^s(x) &=&  \frac{3}{2\sqrt{6} }
 f_\rho^T   (1-2x)  \left[1+ 0.76 (10 x^2 -10 x +1) \right] .
\eeq
 The Gegenbauer polynomial is defined by
 \beq
 C_2^{3/2} (t) = \frac{3}{2} \left (5t^2-1 \right ).
 \eeq

For $\eta$ meson's wave function, $\phi_{\eta_{d\bar{d}}}^A$,
$\phi_{\eta_{d\bar{d}}}^P$ and $\phi_{\eta_{d\bar{d}}}^T$
represent the axial vector, pseudoscalar and tensor components of
the wave function respectively, for which we utilize the result
 from the light-cone sum rule~\cite{ball} including twist-3
contribution:
\beq
\phi_{\eta_{d\bar{d}}}^A(x)&=&\frac{3}{\sqrt{2N_c}}f_xx(1-x)
\left\{ 1+a_2^{\eta_{d\bar{d}}}\frac{3}{2}\left [5(1-2x)^2-1 \right ]\right. \non
&&\left. + a_4^{\eta_{d\bar{d}}}\frac{15}{8}
\left [21(1-2x)^4-14(1-2x)^2+1 \right ]\right \},  \non
\phi^P_{\eta_{d\bar{d}}}(x)&=&\frac{1}{2\sqrt{2N_c}}f_x
\left \{ 1+ \frac{1}{2}\left (30\eta_3-\frac{5}{2}\rho^2_{\eta_{d\bar{d}}} \right )
\left [ 3(1-2x)^2-1 \right] \right.  \non
&& \left. + \frac{1}{8}\left (-3\eta_3\omega_3-\frac{27}{20}\rho^2_{\eta_{d\bar{d}}}-
\frac{81}{10}\rho^2_{\eta_{d\bar{d}}}a_2^{\eta_{d\bar{d}}} \right )
\left [ 35 (1-2x)^4-30(1-2x)^2+3 \right ] \right\} ,  \non
\phi^T_{\eta_{d\bar{d}}}(x)
&=&\frac{3}{\sqrt{2N_c}}f_x(1-2x) \non
 && \cdot \left [ \frac{1}{6}+(5\eta_3-\frac{1}{2}\eta_3\omega_3-
\frac{7}{20}\rho_{\eta_{d\bar{d}}}^2
-\frac{3}{5}\rho^2_{\eta_{d\bar{d}}}a_2^{\eta_{d\bar{d}}})(10x^2-10x+1)\right ],  \non
\eeq
with
\beq
a^{\eta_{d\bar{d}}}_2&=& 0.44, \quad a^{\eta_{d\bar{d}}}_4=0.25,\quad
 a_1^K=0.20, \quad a_2^K=0.25, \non
\rho_{\eta_{d\bar{d}}}&=&m_{\pi}/{m_0^{\eta_{d\bar{d}}}}, \quad
\eta_3=0.015, \quad \omega_3=-3.0.
\eeq

We assume that the wave function of $u\bar{u}$ is same as the wave function of $d\bar{d}$.
For the wave function of the $s\bar{s}$ components, we also use
the same form as $d\bar{d}$ but with $m^{s\bar{s}}_0$ and $f_y$
instead of $m^{d\bar{d}}_0$ and $f_x$, respectively.
For $f_x$ and $f_y$, we use the values as given in Ref.~\cite{kf}
where isospin symmetry is assumed for $f_x$ and $SU(3)$ breaking
effect is included for $f_y$:
 \beq
 f_x=f_{\pi}, \ \ \ f_y=\sqrt{2f_K^2-f_{\pi}^2}.\ \ \
\label{eq:7-5}
\eeq

These values are translated to the values in the two mixing angle
method, which is often used in vacuum saturation approach as:
\beq
f_8 &=&169  {\rm MeV}, \quad f_1=151  {\rm MeV},  \non \theta_8&=&
-25.9^{\circ} (-18.9^{\circ}), \quad \theta_1=-7.1^{\circ}
(-0.1^{\circ}),
\eeq
where the pseudoscalar mixing angle $\theta_p$ is taken as
$-17^{\circ}$ ($-10^{\circ}$) \cite{ekou01}.
The parameters $m_0^i$ $(i=\eta_{d\bar{d}(u\bar{u})}, \eta_{s\bar{s}})$ are
defined as:
\beq
m_0^{\eta_{d\bar{d}(u\bar{u})}}\equiv m_0^\pi \equiv
\frac{m_{\pi}^2}{(m_u+m_d)}, \qquad m_0^{\eta_{s\bar{s}}}\equiv
\frac{2M_K^2-m_{\pi}^2}{(2m_s)}.
 \label{eq:19}
\eeq

We include full expression of twist$-3$ wave functions for light mesons.
The twist$-3$ wave functions are also adopted from QCD sum rule
calculations~\cite{bf}. We will see later that this set of
parameters will give good results for $B \to \rho  \eta^{(\prime)}$ decays.
Using the above chosen wave functions and the central values of relevant
input parameters, we find the numerical values of the corresponding form factors
at zero momentum transfer from Eqs.(\ref{eq:ab}) and (\ref{ls})
\beq
A_0^{B\to \rho}(q^2=0)&=& 0.37, \non
F_0^{B \to \eta}(q^2=0)&=& 0.15, \non
F_0^{B \to \eta^{\prime}}(q^2=0)&=& 0.14. \label{eq:aff0}
\eeq
These values agree well with those as given in Refs.~\cite{ball,kf,sum}.

\subsection{Branching ratios}

For $B \to \rho \etap$ decays, the decay amplitudes in Eqs.~(\ref{eq:m1})
and (\ref{eq:m2}) can be rewritten as
 \beq
{\cal M} &=& V_{ub}^*V_{ud} T -V_{tb}^* V_{td} P= V_{ub}^*V_{ud} T
\left [ 1 + z e^{ i ( \alpha + \delta ) } \right],
\label{eq:ma}
\eeq
where
\beq
z=\left|\frac{V_{tb}^* V_{td}}{ V_{ub}^*V_{ud} } \right|
\left|\frac{P}{T}\right|
\label{eq:zz}
\eeq
is the ratio of penguin to tree contributions,
$\alpha = \arg \left[-\frac{V_{td}V_{tb}^*}{V_{ud}V_{ub}^*}\right]$ is the weak
phase (one of the three CKM angles), and $\delta$ is the relative strong phase
between tree (T) and penguin (P) diagrams.
The ratio $z$ and the strong phase $\delta$ can be calculated in our PQCD
approach. One can leave the CKM angle $\alpha$ as a free
parameter and explore the CP asymmetry parameter dependence on it.

For $B \to \rho^+ \eta$ decay, for example, one can find ``T" and
``P" terms by comparing the decay amplitude as defined in
Eq.~(\ref{eq:m1}) with that in Eq.~(\ref{eq:ma}),
\beq
T(\rho^+\eta)&=& \frac{F_1(\theta_p)}{\sqrt{3}} \cdot \left \{
F_{e\rho} \left( C_1 + \frac{1}{3}C_2\right) f_\eta^d +M_{e\rho}
C_2 \right. \non && \left.+  F_e \left ( \frac{1}{3}C_1 +
C_2\right )
+ \left ( M_a + M_e +M_{a\rho}\right ) C_1 \right \}, \label{eq:tt}\\
P(\rho^+\eta)&=&
\frac{F_1(\theta_p)}{\sqrt{3}} \cdot \left \{
F_{e\rho} \left( -\frac{1}{3}C_3- C_4
-\frac{3}{2}C_7-\frac{1}{2}C_8+\frac{5}{3}C_9+C_{10} \right )
f_\eta^d \right. \non & &\left.
          + F_{e\rho}^{P} \left (\frac{1}{3}C_5+C_6
-\frac{1}{6}C_7-\frac{1}{2}C_{8}\right) \right.\non
&& \left.
+ M_{e\rho} \left(- C_3 + 2 C_6-\frac{3}{2}C_8
+\frac{1}{2}C_9+\frac{3}{2}C_{10}\right) \right.\non
&& \left.
+ F_e \left(\frac{1}{3}C_3+ C_4 +\frac{1}{3}C_9 + C_{10} \right) \right. \non
&& \left.
+ \left ( M_{a\rho}+M_e+M_a \right ) \left(
C_3+C_9 \right) + \left (2M_a^{P}+M_e^{P} \right ) \left ( C_5+
C_7 \right ) \right\}   \non
 &&
 + \frac{F_2(\theta_p)}{\sqrt{3}} \left \{ F_{e\rho}
\left(\frac{1}{2}C_7+\frac{1}{6}C_8-\frac{1}{2}C_9-\frac{1}{6}C_{10}\right)
\cdot f_\eta^s \right. \non
&& \left. + M_{e\rho} \left( C_4
+ C_6 - \frac{1}{2} C_8- \frac{1}{2} C_{10} \right ) \right \}.
\label{eq:pp}
\eeq
Similarly, one can obtain the expressions of
the corresponding tree and penguin terms for the remaining three
decays.

Using the ``T" and ``P" terms, one can calculate the ratio $z$ and
the strong phase $\delta$ for the decay in study. For  $B^+ \to
\rho^+ \eta$ and $\rho^+ \eta'$ decays, we find numerically that
\beq
z(\rho^+\eta) &=&0.10 , \qquad \delta (\rho^+\eta)=-137^\circ , \label{eq:zd1}\\
z(\rho^+\eta') &=& 0.15 , \qquad
\delta(\rho^+\eta')=-139^\circ .\label{eq:zd2}
\eeq
The errors of the ratio $z$ and the strong phase $\delta$  induced by the
uncertainty of the input parameters, such as $\omega_b=0.4 \pm 0.04$ GeV, $m_0^\pi=1.4 \pm
0.1$ GeV, and $\alpha =100^\circ \pm 20^\circ$, are very small in magnitude and not be shown
explicitly in Eqs.~(\ref{eq:zd1}) and (\ref{eq:zd2}).
The reason is that the errors induced by the uncertainties of these input parameters are
canceled almost completely in the ratio.

Unlike the case of QCD factorization approach, the energy scale $t$ (in PQCD factorization
approach ) appeared in the Wilson
coefficients $C_i(t)$ and in the Sudakov form factors $S_j(t)$ vary simultaneously during the
integration over $x_i$  and $b_i$ ($i=1,2,3$). If we choose the hard energy scale $t_j's$ as defined in
Eqs.~(\ref{tf}) with $a_t=1$, there will be no remaining scale dependence
left explicitly after the integration. But we know that such scale dependence should exist and
likely dominate the errors on theoretical predictions for those observables.
Since the calculation in this paper is performed at the leading order and thus may suffers from the
uncertainties due to the scale dependence of the LO Wilson coefficients.
In Ref.~\cite{next} the authors calculated the branching ratios of $B \to K \pi, \pi\pi$ firstly  at the
next-to-leading order by using the PQCD factorization approach, and they found that
the NLO contribution can give about $15-20\%$ correction to LO predictions.
The size of NLO contribution in PQCD approach is indeed very complicated
to calculate. To explore it, as shown in Eq.~(\ref{tf}), we here multiply a factor $a_t= 1\pm 0.1$ to
the ordinary definition of scale $t_j$'s in Refs.~\cite{luy01,kls01,li01,kklls04,wang06},
and take it as an estimation for the uncertainty of the possible scale dependence.
Numerically  we find that
\beq
z(\rho^+ \eta)= 0.10 ^{+0.06}_{-0.01},  \quad \delta (\rho^+ \eta)= (-137 ^{+22}_{-2})^\circ, \\
z(\rho^+ \eta')=0.15 ^{+0.06}_{-0.01}, \quad \delta (\rho^+ \eta')= (-139 ^{+27}_{-1})^\circ,
\eeq
for $a_t=1\pm 0.1$. The larger change of $z$ and $\delta$ corresponds to the case of $a_t=0.9$,
while the magnitude of the variations is consistent with the general expectation.

From Eq.~(\ref{eq:ma}), it is easy to write the decay amplitude
for the corresponding charge conjugated decay mode \beq
\overline{\cal M} &=& V_{ub}V_{ud}^* T -V_{tb} V_{td}^* P =
V_{ub}V_{ud}^* T \left[1 +z e^{i(-\alpha + \delta)} \right].
\label{eq:mb}
 \eeq
Therefore the CP-averaged branching ratio for $B^0 \to \rho \etap$
is
\beq
Br = (|{\cal M}|^2 +|\overline{\cal M}|^2)/2 =  \left|
V_{ub}V_{ud}^* T \right| ^2 \left[1 +2 z\cos \alpha \cos \delta
+z^2 \right], \label{br}
\eeq
where the ratio $z$ and the strong phase $\delta$ have been defined
in Eqs.(\ref{eq:ma}) and
(\ref{eq:zz}). It is easy to see that the CP-averaged branching
ratio is a function of $\cos \alpha$ for the given ratio $z$ and the strong phase
$\delta$. This gives a potential
method to determine the CKM angle $\alpha$ by measuring only the
CP-averaged branching ratios with PQCD calculations. But one should know that
the uncertainty of theory is so large as to make it unrealistic.

Using  the wave functions and the input parameters as specified in
previous sections,  it is straightforward  to calculate the
branching ratios for the four considered decays. The theoretical
predictions in the PQCD approach for the branching ratios of the
decays under consideration are the following
\beq
 Br(\ B^+ \to \rho^+ \eta) &=& \left [8.5 ^{+3.0}_{-2.1}(\omega_b)
 ^{+0.8}_{-0.7}(m_0^\pi) \pm 0.4 (\alpha ) ^{+1.2}_{-0.2}(a_t)\right ] \times 10^{-6},
 \label{eq:brp-eta}\\
 Br(\ B^+ \to \rho^+ \eta^{\prime}) &=&
\left [8.7^{+3.0}_{-2.2}( \omega_b) ^{+0.7}_{-0.9}(m_0^\pi)
^{+0.5}_{-0.7} (\alpha ) ^{+1.1}_{-0.3}(a_t) \right ] \times 10^{-6},
\label{eq:brp-etap}\\
Br(\ B^0 \to \rho^0 \eta) &=& \left [0.024
^{+0.012}_{-0.007}(\omega_b)  ^{+0.004}_{-0.002}(m_0^\pi) \pm
0.002 (\alpha) ^{+0.102}_{-0.005}(a_t) \right ]\times 10^{-6},
\label{eq:br0-eta} \\
Br(\ B^0 \to \rho^0 \eta^{\prime}) &=& \left [0.061
^{+0.030}_{-0.018}(\omega_b)  ^{+0.004}_{-0.003}(m_0^\pi) \pm
0.003 (\alpha ) ^{+0.114}_{-0.009}(a_t)  \right ]\times 10^{-6} \label{eq:br0-etap},
\eeq
for $\theta_p=-10^\circ$; and
\beq
 Br(\ B^+ \to \rho^+ \eta) &=&\left [10.6 ^{+3.9}_{-2.6}(\omega_b)
^{+1.0}_{-0.9}(m_0^\pi)  \pm 0.5 (\alpha ) ^{+1.4}_{-0.3}(a_t) \right ]\times 10^{-6}, \label{eq:brp-eta1}\\
 Br(\ B^+ \to \rho^+ \eta^{\prime}) &=& \left [6.5 ^{+2.3}_{-1.8}( \omega_b)
\pm 0.6(m_0^\pi) \pm 0.5(\alpha ) ^{+0.9}_{-0.2}(a_t) \right ]\times 10^{-6},
\label{eq:brp-etap1} \\
Br(\ B^0 \to \rho^0 \eta) &=&\left [0.042
^{+0.020}_{-0.012}(\omega_b) \pm 0.005(m_0^\pi)
^{+0.006}_{-0.004} ( \alpha ) ^{+0.128}_{-0.012}(a_t) \right ])\times 10^{-6}, \label{eq:br0-eta1}\\
 Br(\ B^0 \to \rho^0 \eta^{\prime}) &=&\left [ 0.047
^{+0.020}_{-0.016}(\omega_b)  ^{+0.001}_{-0.006}(m_0^\pi)
 \pm 0.001 (\alpha ) ^{+0.100}_{-0.004}(a_t) \right ] \times 10^{-6}, \label{eq:br0-etap1}
\eeq
for $\theta_p=-17^\circ$. The major errors are induced by the uncertainty of hard energy scale $t$,
$\omega_b=0.4 \pm 0.04$ GeV, $m_0^\pi = 1.4 \pm 0.1$ GeV and $\alpha =100^\circ \pm
20^\circ$, respectively.
It is easy to see that (a) the errors of the branching ratios induced by varying $a_t$ in the range of
$a_t=[0.9, 1.1]$ are less than $20\%$ for the tree-dominated $B \to \rho^+ \etap$ decays;
but can be significant for the penguin-dominated $B \to \rho^0 \etap$ decays; and
(b) the variations with respect to the central values are large (small) for the case of $a_t=0.9$
($a_t=1.1$). This feature agrees with general expectations:
when the scale $t$ become smaller, the reliability of the
perturbative calculation of the form factors in PQCD approach will become weak!

It is easy to see that the PQCD predictions for the branching
ratios of considered decays agree very well with the measured
values or the upper limits as shown in Eqs.(\ref{eq:exp}) and
(\ref{eq:ulimits}). For the four $B \to \rho \etap$ decays, the
theoretical predictions for the CP-averaged branching ratios in  the PQCD
approach are well consistent with
those given in the QCD factorization approach \cite{bn03b}:
\beq
Br(\ B^+ \to \rho^+ \eta) &=& \left (
9.4^{+5.9}_{-4.8}\right ) \times 10^{-6}, \non
Br(\ B^+ \to \rho^+ \eta^{\prime}) &=& \left (
6.3^{+4.0}_{-3.3}\right ) \times 10^{-6},  \label{eq:br23} \\
Br(\ B^0 \to \rho^0 \eta) &=& \left (
0.03^{+0.17}_{-0.10}\right ) \times 10^{-6}, \non
Br(\ B^0 \to \rho^0 \eta^{\prime}) &=& \left (
0.01^{+0.12}_{-0.06}\right ) \times 10^{-6}, \label{eq:br24}
\eeq
where the individual errors have been added in quadrature.

\begin{figure}[tb]
\centerline{\mbox{\epsfxsize=8cm\epsffile{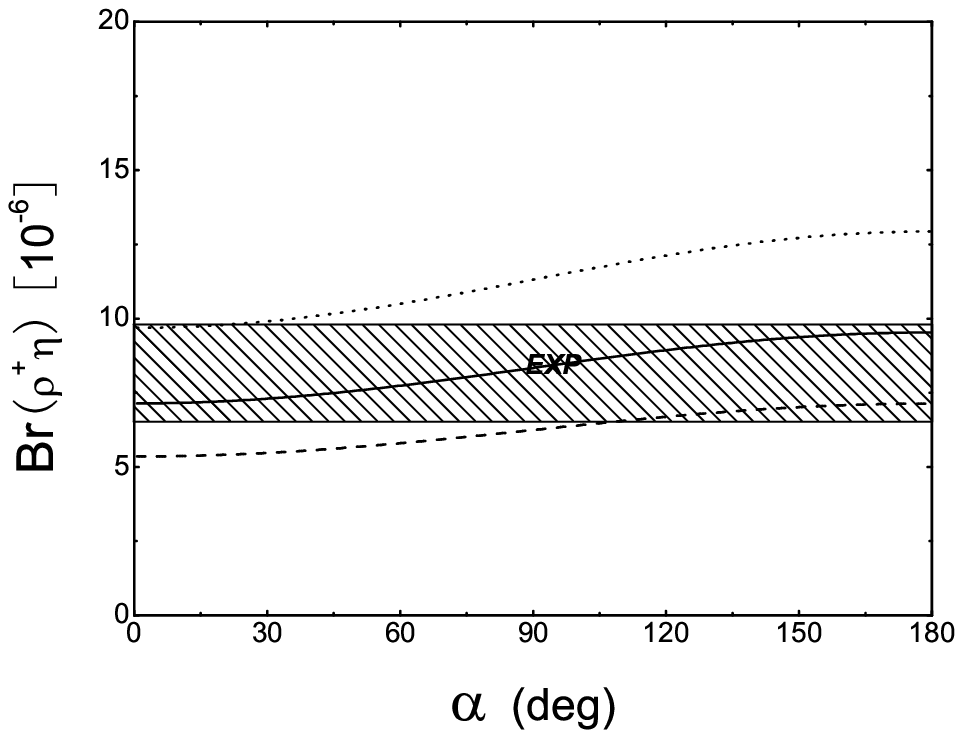}\epsfxsize=8cm\epsffile{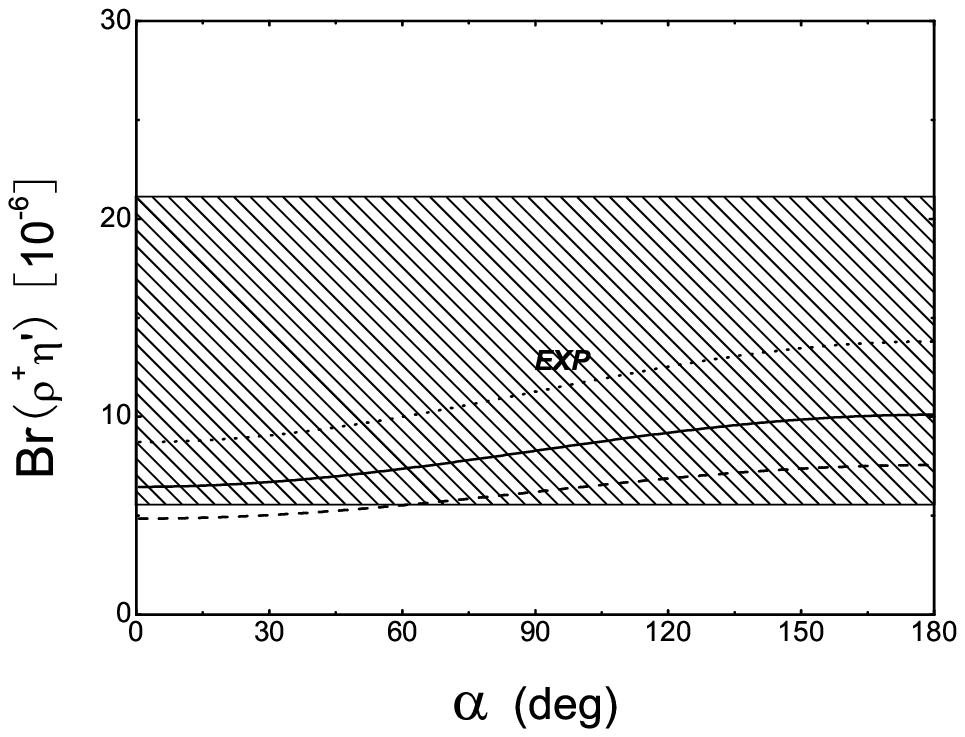}}}
\vspace{0.3cm}
 \caption{The $\alpha$ dependence of the branching ratios (in unit of $10^{-6}$)
 of $B^+\to \rho^+ \etap$ decay for $m_0^\pi=1.4$ GeV,
 $\theta_p=-10^\circ$, $\omega_b=0.36 $ GeV
 (dotted curve), $0.40$ GeV(solid  curve) and $0.44$ GeV(short-dashed curve).
 The gray band show the data.}
 \label{fig:fig3}
\end{figure}

\begin{figure}[tb]
\centerline{\mbox{\epsfxsize=8cm\epsffile{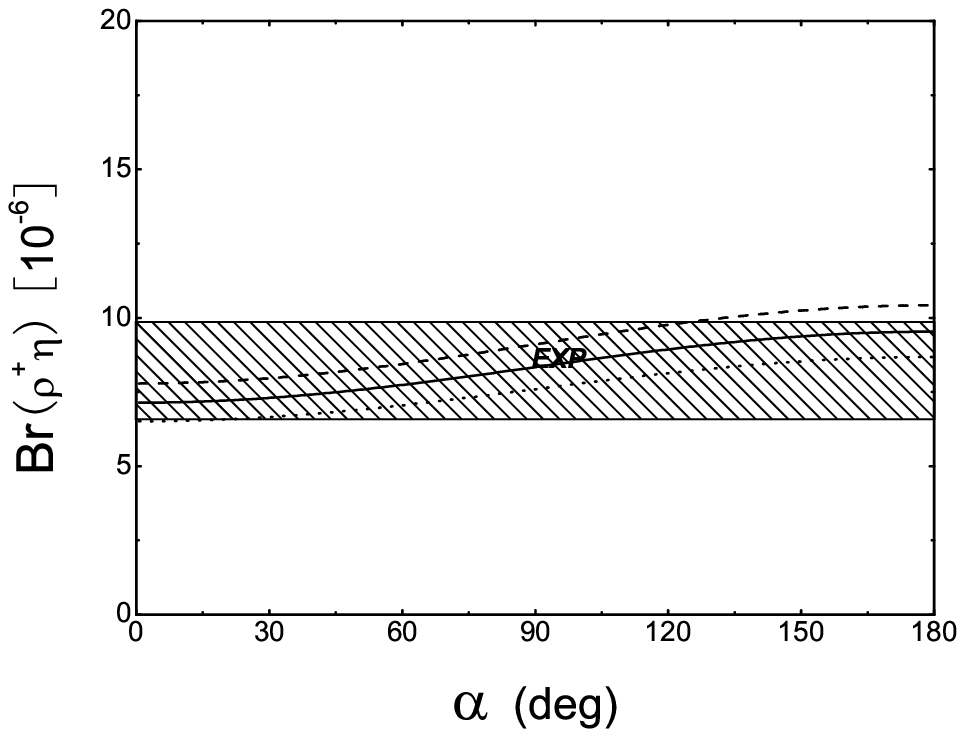}\epsfxsize=8cm\epsffile{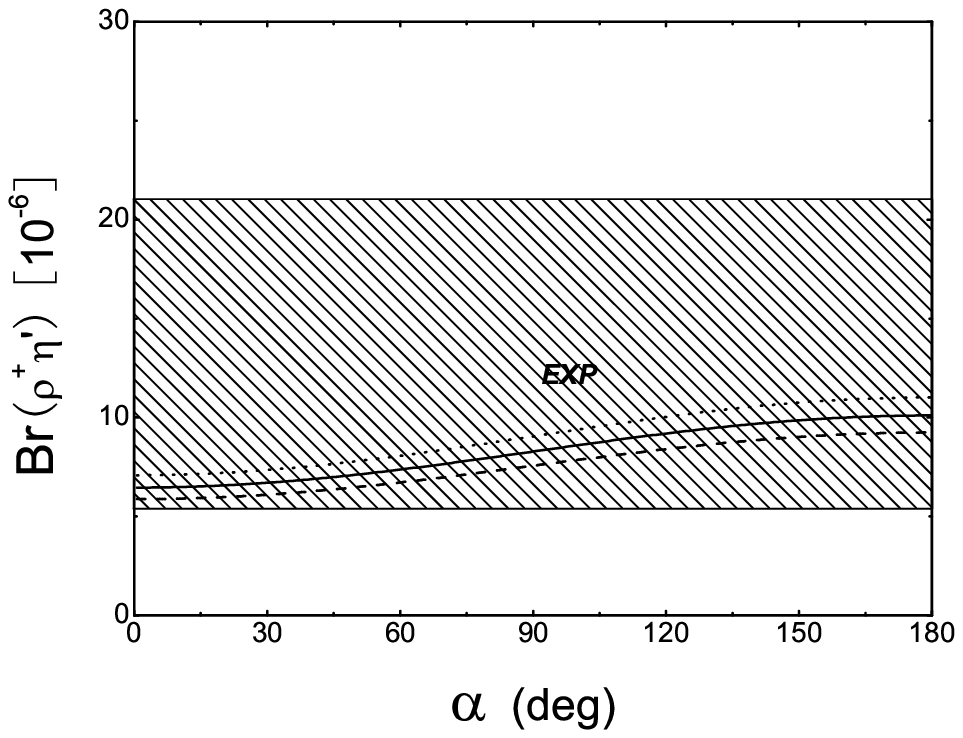}}}
\vspace{0.3cm}
 \caption{The $\alpha$ dependence of the branching ratios (in unit of $10^{-6}$)
 of $B^+\to \rho^+ \etap$ decays for $\omega_b=0.4$ GeV,
$\theta_p=-10^\circ$, $m_0^\pi=1.3$ GeV (dotted curve), $1.4$ GeV
(solid  curve) and $1.5$ GeV (short-dashed curve). The gray band shows the data.}
 \label{fig:fig4}
\end{figure}

\begin{figure}[tb]
\centerline{\mbox{\epsfxsize=8cm\epsffile{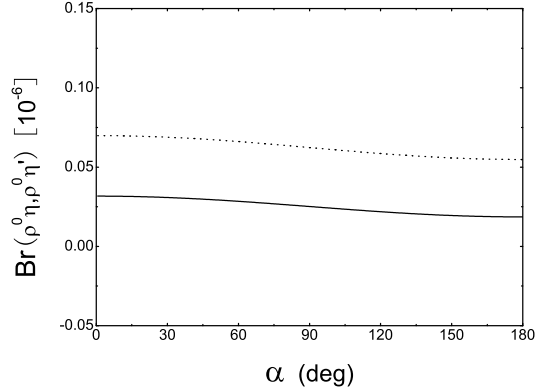}}}
 \caption{The $\alpha$ dependence of the branching ratios (in unit of $10^{-6}$)
 of  $\rho^0 \eta$ (solid  curve) and $\rho^0 \eta^{\prime}$ (dotted curve)  decays
  for $m_0^\pi=1.4$ GeV,$\theta_p=-10^\circ$, $\omega_b=0.40$ GeV .}
 \label{fig:fig5}
\end{figure}

It is worth stressing  that the theoretical predictions in the PQCD approach
have large theoretical errors induced by our ignorance of NLO contributions, and the still large
uncertainties of many input parameters.
In our analysis, we consider the constraints on these
parameters from analysis of other well measured decay channels.
For example, the constraint $1.1 \mbox{GeV} \leq m_0^\pi \leq 1.9 \mbox{GeV}$
was obtained from the phenomenological studies for $B \to \pi \pi$ decays
\cite{luy01}, while the constraint  of $\alpha \approx 100^\circ \pm 20 ^\circ$
were obtained by direct measurements or from the global fit \cite{hfag,charles05,utfit}.
From numerical calculations, we get to know that the main
errors come from the uncertainty of $\omega_b$, $m_0^\pi$,
$\alpha$, $\theta_p$ and the next-to-leading order contributions.

In Figs.~\ref{fig:fig3} and \ref{fig:fig4}, we present,
respectively, the PQCD predictions of the branching ratios of $B
\to \rho^+ \eta$ and $\rho^+ \eta^\prime$ decays for
$\theta_p=10^\circ$, $\omega_b=0.4\pm 0.04$ GeV, $m_0^\pi=1.4\pm
0.1$ GeV and $\alpha=[0^\circ,180^\circ]$. Fig.~\ref{fig:fig5}
shows  the $\alpha$-dependence of the PQCD predictions of the
branching ratios of $B \to \rho^0 \etap$ decays for
$\theta_p=10^\circ$, $\omega_b=0.4$ GeV, $m_0^\pi=1.4$ GeV and
$\alpha=[0^\circ,180^\circ]$.

 From the numerical results and the figures we observe that the
PQCD predictions are very sensitive to the variations of
$\omega_b$ and $m_0^\pi$. The parameter $m_0^\pi$ originates from
the chiral perturbation theory and have a value near 1 GeV. The
$m_0^{\pi}$ parameter characterizes the relative size of twist 3
contribution to twist 2 contribution. Because of the chiral
enhancement of $m_0^{\pi}$, the twist 3 contribution become
comparable in size with the twist 2 contribution. The branching
ratios of $Br(B \to \rho \etap)$ are also sensitive to the
parameter $m_0^\pi$, but not as strong as the $\omega_b$
dependence.

\subsection{CP-violating asymmetries }

Now we turn to the evaluations of the CP-violating asymmetries of
$B \to \rho \etap$ decays in PQCD approach. For $B^+ \to \rho^+
\eta$ and $B^+ \to \rho^+ \eta^\prime$ decays, the direct
CP-violating asymmetries $A_{CP}$ can be defined as:
 \beq
{\cal A}_{CP}^{dir} =  \frac{|\overline{\cal M}|^2 - |{\cal M}|^2 }{
 |\overline{\cal M}|^2+|{\cal M}|^2}=
\frac{2 z \sin \alpha \sin\delta}{1+2 z\cos \alpha \cos \delta
+z^2}, \label{eq:acp1}
 \eeq
where the ratio $z$ and the strong phase $\delta$ have been
defined in previous subsection and are calculable in PQCD
approach.

\begin{figure}[tb]
\vspace{-1cm} \centerline{\epsfxsize=10cm \epsffile{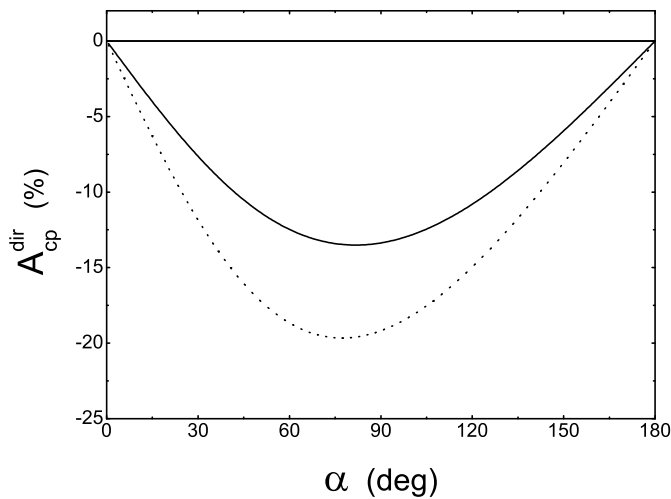}}
\vspace{-0.5cm}
 \caption{The direct CP asymmetries (in percentage) of $B^+\to \rho^+ \eta$ (solid curve)
 and $B^+\to \rho^+ \eta^{\prime}$ (dotted curve) as a function of CKM
angle $\alpha$.}
 \label{fig:fig6}
\end{figure}

Using the central values of $z$ and $\delta$ as given in Eqs.(\ref{eq:zd1}) and
(\ref{eq:zd2}), it is easy  to calculate the CP-violating asymmetries.
In Fig.~\ref{fig:fig6}, we show the $\alpha-$dependence of the direct
CP-violating asymmetries ${\cal A}_{CP}^{dir}$ for $B^\pm \to \rho^\pm \eta$
(the solid curve) and $B^\pm \to \rho^\pm \eta'$ (the dotted curve) decay,
respectively. From Fig.~\ref{fig:fig6}, one can see that the
CP-violating asymmetries $A_{CP}^{dir}(B^\pm  \to \rho^\pm \eta)$ and
$A_{CP}^{dir}(B^\pm  \to \rho^\pm \eta^\prime)$ are
large in magnitude, about $-15\%$ for $\alpha \sim 100^\circ$. The large CP-violating
asymmetries plus large branching ratios are clearly measurable in the B
factory experiments.

For $\alpha = 100^\circ \pm 20^\circ$,  one can read out the
allowed ranges of $A_{CP}^{dir}$ from Fig.~\ref{fig:fig6} directly
\beq
A_{CP}^{dir}(B^\pm \to \rho^\pm \eta) &=& (-13^{+1.2}_{-0.5}(\alpha) ^{+2}_{-14}(a_t) )
\times 10^{-2} \label{eq:acp-a}, \non
A_{CP}^{dir}(B^\pm \to \rho^\pm \eta^\prime) &=&
(-18^{+3.0}_{-1.6}(\alpha) ^{+1 }_{-14} (a_t)) \times 10^{-2} \label{eq:acp-b}.
\eeq
where the second error comes from $a_t=1.0 \pm 0.1$, it is indeed not very large.
The possible theoretical errors induced by the uncertainties of other
input parameters are all very small, since both $z$ and $\delta$ are stable against the variations of them.

The theoretical predictions for the direct CP-violating
asymmetries $A_{CP}^{dir}(B^\pm \to \rho^\pm \etap)$ in the PQCD
approach are generally larger in size than those obtained by using
the QCD factorization approach \cite{bn03b}
\beq
A_{CP}^{dir}(B^\pm \to \rho^\pm \eta) &=& (-2.4 \pm 6.4)\times
10^{-2} , \non A_{CP}^{dir}(B^\pm \to \rho^\pm \eta^\prime) &=&
(4.1 ^{+10.6}_{-9.9}) \times 10^{-2} \label{eq:acp-2}. \eeq

On the experimental side, the new world-average \cite{hfag} is
\beq A_{CP}(B^\pm \to \rho^\pm \eta)^{exp} = ( -3 \pm 16) \times
10^{-2}, \label{eq:acp-ae} \eeq which is still consistent with the
predictions in both PQCD and QCD factorization approach within the
still large experimental error. More data are clearly needed to
make a reliable judgement.

We now study the CP-violating asymmetries for $B^0 \to \rho^0 \etap$ decays.
For these neutral decay modes, the effects of $B^0-\bar{B}^0$ mixing should be
considered.
For $B^0$ meson decays, we know that $\Delta \Gamma/\Delta m_d \ll 1$
and $\Delta \Gamma/\Gamma \ll 1$.
The CP-violating asymmetry of $B^0(\bar B^0) \to \rho^0
\eta^{(\prime)}$ decay is time dependent and can be defined as
\beq
A_{CP} &\equiv& \frac{\Gamma\left (\overline{B_d^0}(\Delta t) \to f_{CP}\right)
- \Gamma\left( B_d^0(\Delta t) \to f_{CP}\right )}{
\Gamma\left (\overline{B_d^0}(\Delta t) \to f_{CP}\right )
+ \Gamma\left (B_d^0(\Delta t) \to f_{CP}\right ) }\non
&=& A_{CP}^{dir} \cos (\Delta m  \Delta t)
+ A_{CP}^{mix} \sin (\Delta m  \Delta t),
\label{eq:acp-def}
\eeq
where $\Delta m$ is the mass difference between the two $B^0$ mass eigenstates,
$\Delta t =t_{CP}-t_{tag} $ is the time difference between the tagged $B^0$
($\overline{B}^0$) and the accompanying $\overline{B}^0$ ($B^0$) with opposite b
flavor decaying to the final CP-eigenstate $f_{CP}$ at the time $t_{CP}$.
The direct and mixing induced CP-violating
asymmetries $A_{CP}^{dir}$ and $A_{CP}^{mix}$ can be written as
\beq
A_{CP}^{dir}=\frac{ \left | \lambda_{CP}\right |^2 -1 }
{1+|\lambda_{CP}|^2}, \qquad A_{CP}^{mix}=\frac{ 2Im
(\lambda_{CP})}{1+|\lambda_{CP}|^2}, \label{eq:acp-dm}
\eeq
where the CP-violating parameter $\lambda_{CP}$ is
\beq
\lambda_{CP} = \frac{ V_{tb}^*V_{td} \langle \rho^0 \etap |H_{eff}|
\overline B^0\rangle} { V_{tb}V_{td}^* \langle \rho^0 \etap
|H_{eff}| B^0\rangle} = e^{2i\alpha}\frac{ 1+z e^{i(\delta-\alpha)} }{
1+ze^{i(\delta+\alpha)} }.
\label{eq:lambda2}
\eeq
Here the ratio $z$ and the strong phase $\delta$ have been defined previously.
In PQCD approach, since both $z$ and $\delta$ are calculable,
it is easy to find the numerical values of
$A_{CP}^{dir}$ and $A_{CP}^{mix}$ for the considered decay processes.

For $B^0 \to \rho^0 \eta$ and $\rho^0 \eta'$ decays,
the numerical values of the ratio $z$ and the strong phase
$\delta$ are
\beq
z(\rho^0\eta) &=&4.0, \qquad \delta
(\rho^0\eta)=-57^\circ, \label{eq:zd3}\\
z(\rho^0\eta') &=& 6.8, \qquad
\delta(\rho^0\eta')=-65^\circ.\label{eq:zd4}
\eeq
Unlike the case of $B^\pm \to \rho^\pm \etap$ decays, we here have $z > 1$,
which means that the ``P" term is
much larger in size than the ``T" term for $B^0\to \rho^0 \etap$
decays, since the ``T'' term here is a color-suppressed tree.

\begin{figure}[tb]
\vspace{-1cm} \centerline{\epsfxsize=10cm \epsffile{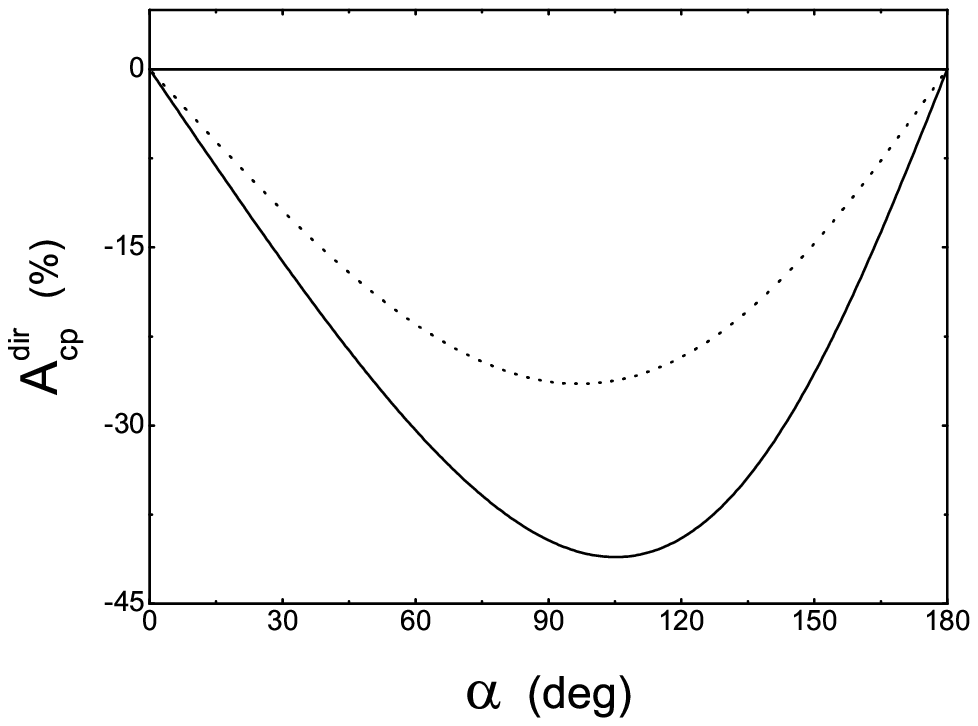}}
\vspace{-0.5cm}
 \caption{The direct CP asymmetry $A_{CP}^{dir}$ (in percentage) of
 $B^0\to \rho^0 \eta$ (solid curve)
 and $B^0\to \rho^0 \eta^{\prime}$ (dotted curve) as a function of CKM
angle $\alpha$.} \label{fig:fig7}
\end{figure}

\begin{figure}[htb]
\vspace{-1cm} \centerline{\epsfxsize=10 cm \epsffile{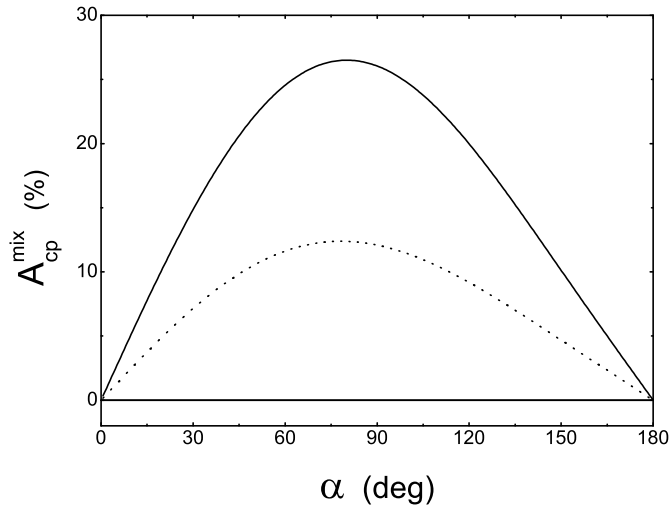}}
\vspace{-0.5cm} \caption{The mixing induced CP asymmetry
$A_{CP}^{mix}$ (in percentage) of $B^0\to \rho^0 \eta$ (solid
curve) and $B^0\to \rho^0 \eta^{\prime}$ (dotted curve) as a
function of CKM angle $\alpha$ .} \label{fig:fig8}
\end{figure}

In Figs.~\ref{fig:fig7} and \ref{fig:fig8}, we show the
$\alpha-$dependence of the direct and the mixing-induced CP-violating asymmetry
$A_{CP}^{dir}$ and $A_{CP}^{mix}$ for $B^0 \to \rho^0 \eta$ (solid curve) and $B^0
\to \rho^0 \eta^\prime$  (dotted curve) decays, respectively. For
$\alpha\sim 100^\circ$, one can find numerically that
\beq
A_{CP}^{dir}(B^0 \to \rho^0 \eta) &\approx & -41 \% , \qquad
A_{CP}^{mix}(B^0 \to \rho^0 \eta) \approx  +25\% , \label{eq:acp3}\\
A_{CP}^{dir}(B^0 \to\rho^0 \eta^\prime)&\approx& -27 \% , \qquad
A_{CP}^{mix}(B^0 \to \rho^0 \eta^\prime) \approx +11\%.
\label{eq:acp-4}
\eeq
They are also large in size. The theoretical errors induced by the uncertainties
of input parameters are only about $10\%$ because of the cancelation in
ratios. If we vary $a_t$ in the range of $0.9 \leq a_t \leq 1.1$, however,
the theoretical predictions for the CP-violating asymmetries of
the penguin-dominated $B^0 \to \rho^0 \etap$ decays may change significantly
\beq
A_{CP}^{dir}(B^0 \to \rho^0 \eta) &= & [-85\%, +24\%], \qquad
A_{CP}^{mix}(B^0 \to \rho^0 \eta) =[-19\%, +35\%], \label{eq:acp5}\\
A_{CP}^{dir}(B^0 \to\rho^0 \eta^\prime)&=&[-75\% , +13\%], \qquad
A_{CP}^{mix}(B^0 \to \rho^0 \eta^\prime) =[ -9 \%, +22\%].
\label{eq:acp-6}
\eeq
This feature may be interpreted as an indication of the importance of the
NLO contributions to those penguin dominated decay modes.

\begin{figure}[htb]
\vspace{-1cm} \centerline{\epsfxsize=10 cm \epsffile{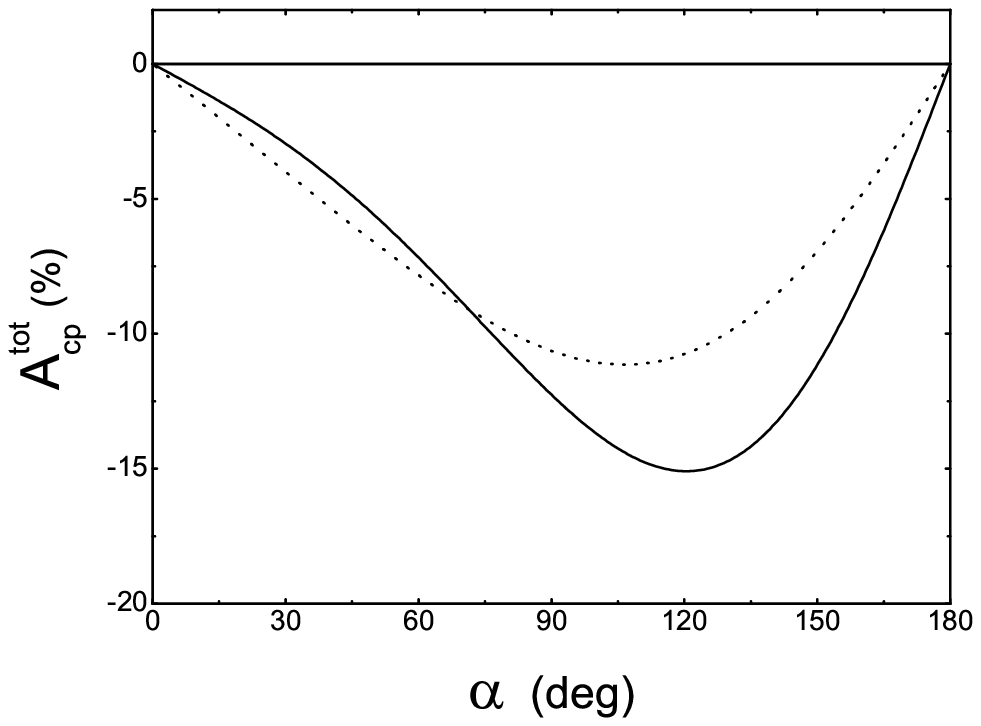}}
\vspace{-0.5cm} \caption{The total CP asymmetry $A_{cp}^{tot}$ (in
percentage) of $B^0\to \rho^0 \eta$ (solid curve) and $B^0\to
\rho^0 \eta^{\prime}$ (dotted curve) as a function of CKM angle
$\alpha$ .} \label{fig:fig9}
\end{figure}

If we integrate the time variable $t$, we will get the total CP
asymmetry for $B^0 \to \rho^0 \etap$ decays,
\beq
A_{CP}=\frac{1}{1+x^2} A_{CP}^{dir} + \frac{x}{1+x^2} A_{CP}^{mix},
\eeq
where $x=\Delta m/\Gamma=0.771$ for the
$B^0-\overline{B}^0$ mixing \cite{pdg04}. In Fig.\ref{fig:fig9},
we show the $\alpha$-dependence of the total CP asymmetry $A_{CP}$
for $B^0 \to \rho^0 \eta$ (solid curve) and $B^0 \to \rho^0
\eta^\prime$ (dotted curve) decay, respectively. For $\alpha\sim 100^\circ$,
the total CP asymmetry is around $-10\%$ for $B^0 \to \rho^0 \etap $ decays.

For $B^+ \to \rho^+ \eta$ and $\rho^+ \eta^\prime$ decays,
the large CP-violating asymmetry around $-15\%$ could be measured in the running
B factory experiments since their branching ratios are rather large,
$\sim 10^{-5}$.
For $Br(B^0 \to \rho^0 \eta)$ and $\rho^0 \eta^\prime$ decays, however,
it is very difficult to measure their CP-violating asymmetries at current running
B factories since their branching ratios are very small, $\sim 10^{-8}$ only.
It could be measured in the forthcoming LHCb experiment.

\subsection{Effects of possible gluonic component of $\eta^\prime$}

Up to now, we have not considered the possible  contributions to the branching
ratios and CP-violating asymmetries of $B \to \rho \eta^\prime$ decays induced by
the possible gluonic component of $\eta^\prime$ \cite{rosner83,ekou01,ekou02}.
When $Z_{\eta^\prime} \neq 0$, a decay amplitude ${\cal M}'$ will be produced
by the gluonic component of $\eta^\prime$. Such decay amplitude may
interfere constructively or destructively with the ones from the $q\bar{q}$ ($q=u,d,s$) components of
$\eta^\prime$, the branching ratios of the decays in question
may be increased or decreased accordingly.

Unfortunately, we currently do not know how to calculate this kind of contributions
reliably. But we can treat it as an theoretical uncertainty.
For $|M'/M(q\bar{q})| \sim 0.1-0.2$, for example, the resulted uncertainty
for the branching ratios as given in Eqs.(\ref{eq:brp-etap}) and (\ref{eq:br0-etap})
will be around twenty to thirty percent.

 From Eq.~(\ref{eq:brp-etap}), one can see that the theoretical
prediction of $Br(B^+ \to \rho^+ \eta^\prime)$ in the PQCD
approach agrees well with the measured value within one standard
deviation, which means that there is no large room left for the
contribution due to the gluonic component of $\eta^\prime$. We
therefore believe that the gluonic admixture of $\eta^\prime$
should be small, and most possibly not as important as expected
before.

As for the CP-violating asymmetries of $B \to \rho \eta^\prime$ decays,
the possible contributions of the gluonic components of the $\eta^\prime$ meson are
largely cancelled in the ratio.

\section{summary }

In this paper,  we calculate the branching ratios and CP-violating asymmetries
of $B^0 \to \rho^0 \eta$, $B^0 \to \rho^0 \eta^{\prime}$,
$B^+ \to \rho^+ \eta$, and $B^+ \to \rho^+ \eta^{\prime}$ decays in the
PQCD factorization approach.

Besides the usual factorizable diagrams, the non-factorizable and annihilation
diagrams as shown in Figs.~(\ref{fig:fig1}) and (\ref{fig:fig2}) are also
calculated analytically. Although the non-factorizable and annihilation contributions
are sub-leading for the branching ratios of the considered decays,
but they are not negligible. Furthermore these diagrams provide the
necessary strong phase required by a non-zero CP-violating asymmetry for the
considered decays.

 From our calculations and phenomenological analysis, we found the following
results:
\begin{itemize}

\item From analytical calculations, the form factors for $B \to
\eta$, $B \to \eta^\prime$ and $B\to \rho$ transitions can be
extracted. The PQCD results for these form factors are $A_0^{B\to
\rho}(0)=0.37$, $F_0^{B\to \eta}(0)=0.15$ and $F_0^{B\to
\eta^\prime}(0)=0.14$, which are in good agreement with those
obtained from the QCD sum rule calculations.

\item
For the branching ratios of the four considered decay modes, the theoretical
predictions in PQCD approach are
\beq
Br(B^+ \to \rho^+ \eta^{(\prime)}) \approx 9 \times 10^{-6}, \\
Br(B^0 \to \rho^0 \eta^{(\prime)}) \approx 5 \times 10^{-8}.
\eeq
Although the theoretical uncertainties are still large (can reach
$50\%$), the leading PQCD predictions for the branching ratios
agree well with the measured values or currently available
experimental upper limits, and are also consistent with the
results obtained by employing the QCD factorization approach.

\item
For the CP-violating asymmetries, the theoretical predictions in PQCD approach are
\beq
A_{CP}^{dir}(B^\pm \to \rho^\pm \eta) & \approx & -13\% , \\
A_{CP}^{dir}(B^\pm \to \rho^\pm \eta^{\prime}) & \approx & -18\% , \\
A_{CP}^{dir}(B^0 \to \rho^0 \eta) & \approx & -41\%, \quad
A_{CP}^{mix}(B^0 \to \rho^0 \eta)  \approx  +25\%, \\
A_{CP}^{dir}(B^0 \to \rho^0 \eta^{\prime}) & \approx & -27\%, \quad
A_{CP}^{mix}(B^0 \to \rho^0 \eta^\prime)  \approx  +11\%,
\eeq
for $\alpha \approx 100^\circ$. For $B^\pm \to \rho^\pm \eta$
decay, the CP-violating asymmetry around $-15\%$ could be measured
in the running B factory experiments. For the neutral decays, their
CP-violating asymmetries may be measured in the forthcoming
LHCb experiments.

\item
The major theoretical errors of the computed observables are induced by the uncertainties of the
hard energy scale $t_j$'s, the parameters $\omega_b$ and $m_0^\pi$, as well as the CKM  angle $\alpha$.

\item
 From the good consistency of the PQCD prediction of  $Br(B^+ \to \rho^+ \eta^\prime)$
with the measured value, we believe that the gluonic admixture of $\eta^\prime$
should be small, and most possibly not as important as expected before.

\end{itemize}

\begin{acknowledgments}

We are very grateful to Li Ying for helpful discussions.
This work is partly supported  by the National
Natural Science Foundation of China under Grant
No.90103013, 10135060,10275035,10475085,10575052,
and by the Research Foundations of Jiangsu Education Committee and
Nanjing Normal University under Grant No.~214080A916 and 2003102TSJB137.

\end{acknowledgments}


\begin{appendix}

\section{Related Functions }\label{sec:aa}

We show here the function $h_i$'s, coming from the Fourier
transformations  of $H^{(0)}$,
 \beq
 h_e(x_1,x_3,b_1,b_3)&=&
 K_{0}\left(\sqrt{x_1 x_3} m_B b_1\right)
 \left[\theta(b_1-b_3)K_0\left(\sqrt{x_3} m_B
b_1\right)I_0\left(\sqrt{x_3} m_B b_3\right)\right.
 \non
& &\;\left. +\theta(b_3-b_1)K_0\left(\sqrt{x_3}  m_B b_3\right)
I_0\left(\sqrt{x_3}  m_B b_1\right)\right] S_t(x_3), \label{he1}
\eeq
 \beq
 h_a(x_2,x_3,b_2,b_3)&=&
 K_{0}\left(i \sqrt{x_2 x_3} m_B b_3\right)
 \left[\theta(b_3-b_2)K_0\left(i \sqrt{x_3} m_B
b_3\right)I_0\left(i \sqrt{x_3} m_B b_2\right)\right.
 \non
& &\;\;\;\;\left. +\theta(b_2-b_3)K_0\left(i \sqrt{x_3}  m_B
b_2\right) I_0\left(i \sqrt{x_3}  m_B b_3\right)\right] S_t(x_3),
\label{he3}
\eeq
 \beq
 h_{f}(x_1,x_2,x_3,b_1,b_2) &=&
 \biggl\{\theta(b_2-b_1) \mathrm{I}_0(M_B\sqrt{x_1 x_3} b_1)
 \mathrm{K}_0(M_B\sqrt{x_1 x_3} b_2)
 \non
&+ & (b_1 \leftrightarrow b_2) \biggr\}  \cdot\left(
\begin{matrix}
 \mathrm{K}_0(M_B F_{(1)} b_1), & \text{for}\quad F^2_{(1)}>0 \\
 \frac{\pi i}{2} \mathrm{H}_0^{(1)}(M_B\sqrt{|F^2_{(1)}|}\ b_1), &
 \text{for}\quad F^2_{(1)}<0
\end{matrix}\right),
\label{eq:pp1}
 \eeq
 \beq
 h_f^1(x_1,x_2,x_3,b_1,b_2) &=&
 \biggl\{\theta(b_1-b_2) \mathrm{K}_0(i \sqrt{x_2 x_3} b_1 M_B)
 \mathrm{I}_0(i \sqrt{x_2 x_3} b_2 M_B)
 \non
&+& (b_1 \leftrightarrow b_2) \biggr\} \cdot \left(
\begin{matrix}
 \mathrm{K}_0(M_B F_{(2)} b_1), & \text{for}\quad F^2_{(2)}>0 \\
 \frac{\pi i}{2} \mathrm{H}_0^{(1)}(M_B\sqrt{|F^2_{(2)}|}\ b_1), &
 \text{for}\quad F^2_{(2)}<0
\end{matrix}\right),
\label{eq:pp3} \eeq \beq h_f^2(x_1,x_2,x_3,b_1,b_2) &=&
\biggl\{\theta(b_1-b_2) \mathrm{K}_0(i \sqrt{x_2 x_3} b_1 M_B)
 \mathrm{I}_0(i \sqrt{x_2 x_3} b_2 M_B)+(b_1 \leftrightarrow b_2) \biggr\}
 \non
& & \cdot
 \frac{\pi i}{2} \mathrm{H}_0^{(1)}(\sqrt{x_1+x_2+x_3-x_1 x_3-x_2 x_3}\ b_1 M_B),
 \label{eq:pp4}
\eeq
where $J_0$ is the Bessel function and  $K_0$, $I_0$ are
modified Bessel functions $K_0 (-i x) = -(\pi/2) Y_0 (x) + i
(\pi/2) J_0 (x)$, and $F_{(j)}$'s are defined by
\beq
F^2_{(1)}&=&(x_1 -x_2) x_3\;,\\
F^2_{(2)}&=&(x_1-x_2) x_3\;\;.
 \eeq

The threshold resummation form factor $S_t(x_i)$ is adopted from
Ref.\cite{kurimoto} \beq S_t(x)=\frac{2^{1+2c} \Gamma
(3/2+c)}{\sqrt{\pi} \Gamma(1+c)}[x(1-x)]^c, \eeq where the
parameter $c=0.3$. This function is normalized to unity.

The Sudakov factors used in the text are defined as \beq
S_{ab}(t) &=& s\left(x_1 m_B/\sqrt{2}, b_1\right) +s\left(x_3
m_B/\sqrt{2}, b_3\right) +s\left((1-x_3) m_B/\sqrt{2}, b_3\right)
\non
&&-\frac{1}{\beta_1}\left[\ln\frac{\ln(t/\Lambda)}{-\ln(b_1\Lambda)}
+\ln\frac{\ln(t/\Lambda)}{-\ln(b_3\Lambda)}\right],
\label{wp}\\
S_{cd}(t) &=& s\left(x_1 m_B/\sqrt{2}, b_1\right)
 +s\left(x_2 m_B/\sqrt{2}, b_2\right)
+s\left((1-x_2) m_B/\sqrt{2}, b_2\right) \non
 && +s\left(x_3
m_B/\sqrt{2}, b_1\right) +s\left((1-x_3) m_B/\sqrt{2}, b_1\right)
\non
 & &-\frac{1}{\beta_1}\left[2
\ln\frac{\ln(t/\Lambda)}{-\ln(b_1\Lambda)}
+\ln\frac{\ln(t/\Lambda)}{-\ln(b_2\Lambda)}\right],
\label{Sc}\\
S_{ef}(t) &=& s\left(x_1 m_B/\sqrt{2}, b_1\right)
 +s\left(x_2 m_B/\sqrt{2}, b_2\right)
+s\left((1-x_2) m_B/\sqrt{2}, b_2\right) \non
 && +s\left(x_3
m_B/\sqrt{2}, b_2\right) +s\left((1-x_3) m_B/\sqrt{2}, b_2\right)
\non
 &
&-\frac{1}{\beta_1}\left[\ln\frac{\ln(t/\Lambda)}{-\ln(b_1\Lambda)}
+2\ln\frac{\ln(t/\Lambda)}{-\ln(b_2\Lambda)}\right],
\label{Se}\\
S_{gh}(t) &=& s\left(x_2 m_B/\sqrt{2}, b_1\right)
 +s\left(x_3 m_B/\sqrt{2}, b_2\right)
+s\left((1-x_2) m_B/\sqrt{2}, b_1\right) \non
 &+& s\left((1-x_3)
m_B/\sqrt{2}, b_2\right)
-\frac{1}{\beta_1}\left[\ln\frac{\ln(t/\Lambda)}{-\ln(b_1\Lambda)}
+\ln\frac{\ln(t/\Lambda)}{-\ln(b_2\Lambda)}\right], \label{ww}
\eeq
where the function $s(q,b)$ are defined in the Appendix B of Ref.~\cite{luy01} and
the hard energy scale $t_j'$s have been given in Eq.~(\ref{tf}).

\end{appendix}



\newpage

\begin{thebibliography}{99}

\bibitem{bbns}
M.~Beneke, G.~Buchalla, M.~Neubert, and C.T.~Sachrajda, \prl {\bf 83}, 1914 (1999).

\bibitem{bn03b}
M.~Beneke and  M.~Neubert, \npb{\bf 675},  333 (2003).

\bibitem{lb80}
G.P.~Lepage and S.Brodsky, \prd {\bf 22}, 2157 (1980);
J.~Potts and G.~Sterman, \npb {\bf 225}, 62 (1989.

\bibitem{cl97}
C.-H. V.~Chang and H.-n.~Li,\prd {\bf 55}, 5577 (1997);
T.-W.~Yeh and H.-n.~Li, \prd {\bf 56}, 1615 (1997).

\bibitem{li2003}
H.-n.~Li, Prog.Part.$\&$ Nucl.Phys. {\bf 51}, 85 (2003), and reference therein.

\bibitem{ds02}
S.~Descotes-Genon and C.T.Sachrajda, \npb {\bf 625}, 239 (2002);
H.-n. Li, p360-364, Proceedings of International Conference on
Flavor Physics, Zhang-Jia-Jie 2001, Flavor physics,
hep-ph/0110365; Z.-T. Wei, M.-Z. Yang, Nucl. Phys. B642, 263-289
(2002);  M.~Beneke and T.~feldmann, \npb {\bf 685}, 296 (2004).

\bibitem{bene}
M.~Beneke, G.~Buchalla, M.~Neubert, and C.T.~Sachrajda,\npb {\bf 591}, 313 (2000).

\bibitem{du02}
D.S.~Du, H.J.~Gong, J.F.~Sun, D.S.~Yang, and G.H.~Zhu, \prd {\bf 65}, 094025 (2002).

\bibitem{luy01}
C.-D.~L\"u, K.~Ukai and M.-Z.~Yang,\prd {\bf 63}, 074009 (2001).

\bibitem{kls01}
Y.-Y.~Keum, H.-n.~Li and A.I.~Sanda,\plb 504, {\bf 6} (2001);
  \prd {\bf 63}, 054008 (2001).

\bibitem{li01}
H.-n. Li, \prd D64, 014019 (2001); S.~Mishima,\plb {\bf 521}, 252 (2001);
C.-H.~Chen, Y.-Y.~Keum, and H.-n.~Li,\prd {\bf 64}, 112002 (2001);
A.I.~Sanda and K.~Ukai, Prog. Theor. Phys. {\bf 107}, 421 (2002);
C.D.~L\"u,\epjc {\bf 24}, 121 (2002);
C.-H.~Chen,Y.-Y.~Keum, and H.-n.~Li,\prd {\bf 66}, 054013 (2002);
Y.-Y.~Keum and A.I.~Sanda, \prd {\bf 67}, 054009(2003).

\bibitem{kklls04}
Y.-Y.~Keum,  T.~Kurimoto, H.-n.~Li, C.D.L\"u, and A.I.~Sanda, \prd {\bf 69}, 094018 (2004);
Y.~Li and C.D.~L\"u, \jpg {\bf 29}, 2115 (2003);
C.D.~L\"u, \prd {\bf 68}, 097502 (2003);
Y.~Li, C.D.~L\"u, Z.J.~Xiao, and X.Q.~Yu, \prd {\bf 70}, 034009  (2004);
Y.~Li, C.D.~L\"u, and Z.J.~Xiao, \jpg {\bf 31}, 273 (2005);
X.Q~Yu, Y.~Li and C.D.~L\"u, \prd {\bf 71}, 074026 (2005);
C.D.~L\"u, Y.L.~Shen and J.~Zhu, \epjc 41, 311 (2005);
J.~Zhu, Y.L.~Shen and C.D.~L\"u, \prd 72, 054015 (2005);  and \jpg 32, 101 (2006);
Y.~Li and C.D.~L\"u, Commun.Theor.Phys. {\bf 44}, 659 (2005); and \prd {\bf 73}, 014024 (2006);
C.D.~L\"u, M.~Matsumori, A.I.~Sanda, and M.Z.~Yang, \prd {\bf 72}, 094005 (2005).

\bibitem{wang06}
H.S.~Wang, X.~Liu, Z.J.~Xiao, L.B.~Guo and C.D.~L\"u, \npb {\bf 738}, 243 (2006).

\bibitem{buras96}
G.~Buchalla,A.J.~Buras,M.E.~Lautenbacher, Rev. Mod. Phys. {\bf 68}, 1125 (1996).

\bibitem{ali98}
A.~Ali, G.~Kramer, and C.D.~L\"u, \prd {\bf 58}, 094009 (1998), {\it ibid}
{\bf 59}, 014005 (1999);
Y.-H. Chen, H.-Y. Cheng, B. Tseng, and K.-C. Yang, \prd {\bf 60}, 094014 (1999).


\bibitem{babar}
BaBar Collaboration, B.~Aubert {\it et al.}, \prd {\bf 70}, 032006 (2004);
BaBar Collaboration, B.~Aubert {\it et al.}, \prl {\bf 95}, 131803 (2005).

\bibitem{belle}
Belle Collaboration, P.~Chang {\it et al.}, \prd {\bf 71}, 091106(R) (2005);
K.~Abe{\it et  al.}, BELLE-CONF-0408, ICHEP04 11-0653, Aug. 2004.

\bibitem{cleo}
CLEO Collaboration, A.~Gritsan, {\it et al.},
talk given at Lake Louis Winter Institute, Feb. 20-26, (2000).

\bibitem{hfag}
Heavy Flavor Averaging Group, http://www.slac.stanford.edu/xorg/hfag.

\bibitem{li02}
H.-n.~Li, \prd {\bf 66}, 094010 (2002).

\bibitem{soft}
H.-n.~Li and B.~Tseng,\prd {\bf 57}, 443, (1998).

\bibitem{grozin}
A.G.~Grozin and M.~Neubert, \prd {\bf 55}, 272 (1997);
M.~Beneke amd T.~Feldmann, \npb {\bf 592}, 3 (2001).


\bibitem{kurimoto}
T.~Kurimoto, H.-n.~Li, and A.I.~Sanda, \prd {\bf 65}, 014007 (2001);
C.D.~Lu and M.Z.~Yang, \epjc {\bf 28}, 515 (2003).

\bibitem{cl00}
C.-h.~Chen and H.-n.~Li, \prd {\bf 63}, 014003 (2000).

\bibitem{ball2}
P.~Ball, V.M.~Braun, Y.~Koike, and K.~Tanaka, \npb {\bf 529}, 323 (1998)

\bibitem{ly}
C.D.~L\"u and M.Z.~Yang, \epjc {\bf 23}, 275 (2002)

\bibitem{pdg04}
Particle Data Group, S.~Eidelman {\it et al.}, \plb {\bf 592}, 1 (2004).

\bibitem{ekou01}
E.~Kou, \prd {\bf 63}, 054027 (2001).

\bibitem{ekou02}
E.~Kou and A.I.~Sanda,\plb {\bf 525}, 240 (2002).

\bibitem{rosner83}
J.L.~Rosner, \prd {\bf 27}, 1101 (1983).

\bibitem{ball}
P.~Ball, J. High Energy Phys. 9809, 005 (1998);
{\it ibid }, 9901, 010 (1999).

\bibitem{kf}
T.~Feldmann and P.~Kroll, \epjc {\bf 5}, 327 (1998)

\bibitem{bf}
V.M.~Braun and I.E.~Filyanov , Z. Phys. C {\bf 48}, 239 (1990);
P.~Ball, J. High Energy Phys. 01, 010 (1999).

\bibitem{sum}
 P.~Ball and V.M.~Braun, \prd {\bf 58}, 094016 (1998).

\bibitem{charles05}
J.~Charles {\it et al.}, \epjc {\bf 41}, 1 (2005).

\bibitem{utfit}
UTfit Collaboration, M.~Bona {\it et al.}, JHEP 0507 (2005) 028.

\bibitem{next}
H.-n.~Li, S.~Mishima, A.I.~Sanda, \prd {\bf 72}, 114005 (2005).

\end{thebibliography}
\end{document}